\documentstyle[12pt]{article}
\def\ln{{\rm ln}}

\def\a{\begin{eqnarray}}
\def\b{\end{eqnarray}}
\def\0{\nonumber}
\def\ba{\begin{array}}
\def\ea{\end{array}}
\def\noal{\noalign{\vskip10pt}}

\def\q{{\bar{\cal Q}}}

\def\al{{\alpha}}
\def\lm{{\lambda}}

\def\cm{{\cal M}}

\renewcommand{\theequation}{\thesection.\arabic{equation}}

\newlength{\extraspace}
\newlength{\extraspaces}
\newcounter{dummy}
\newcommand{\ai}{
\addtocounter{equation}{1}
\setcounter{dummy}{\value{equation}}
\setcounter{equation}{0}
\renewcommand{\theequation}{\thesection.\arabic{dummy}\alph{equation}}
\begin{eqnarray}
\addtolength{\abovedisplayskip}{\extraspaces}
\addtolength{\belowdisplayskip}{\extraspaces}
\addtolength{\abovedisplayshortskip}{\extraspace}
\addtolength{\belowdisplayshortskip}{\extraspace}}
\newcommand{\bj}{
\end{eqnarray}
\setcounter{equation}{\value{dummy}}
\renewcommand{\theequation}{\thesection.\arabic{equation}}}
\def\d{{\partial}}

\newcommand{\ddlm}[1]{{\partial \over \partial \lm_{#1}}}

\begin{document}
\begin{flushright}
SISSA-ISAS 172/93/EP\\
BONN-HE-45/93\\
hep-th/9311089
\end{flushright}
\vskip0.5cm
\centerline{\LARGE\bf Correlation functions of two--matrix models}
\vskip0.3cm
\centerline{\large  L.Bonora}
\centerline{International School for Advanced Studies (SISSA/ISAS)}
\centerline{Via Beirut 2, 34014 Trieste, Italy}
\centerline{INFN, Sezione di Trieste.  }
\vskip0.5cm
\centerline{\large C.S.Xiong}
\centerline{Physikalisches Institut der Universit\"at Bonn}
\centerline{Nussallee 12, 53115 Bonn, Germany}
\vskip5cm
\abstract{We show how to calculate correlation functions of two
matrix models. Our method consists in making full use of the integrable
hierarchies and their reductions, which were shown in previous papers
to naturally appear in multi--matrix models. The second ingredient
we use are the $W$--constraints. In fact an explicit solution
of the relevant hierarchy, satisfying the $W$--constraints (string
equation), underlies the  explicit calculation of the correlation
functions. In the course of our derivation we do not use any
continuum limit tecnique. This allows us to find many solutions
which are invisible to the latter technique. }

\vfill\eject

\section{Introduction}

\setcounter{equation}{0}
\setcounter{footnote}{0}

Matrix models are believed to provide a (discrete)
description of two dimensional gravity coupled to
matter. Multi--matrix models
might also provide a new approach to non--perturbative QCD.
An accurate study of these models is therefore strongly motivated.
While one--matrix models have been widely investigated, our knowledge
of multi--matrix models is not as satisfactory (see however \cite{Douglas},
\cite{G},\cite{GN},\cite{is},\cite{ising},\cite{tada},\cite{DEB},
\cite{R},\cite{AS},\cite{MM},
\cite{PR},\cite{DKK},\cite{Stau}).

In this paper we bring to a conclusion an analysis started in previous
papers \cite{BX1},\cite{BX2},\cite{BX3},\cite{BX4} (see also \cite{BMX},
\cite{X1}). For simplicity we concentrate on two--matrix models and show
how to find {\it exact genus by genus solutions} in these models \footnote{
Our approach is not of the type initiated by M.Kontsevitch \cite{Ko},
although in a few cases the two approaches provide equivalent descriptions.}.
By exact solutions
we mean that we calculate the correlation functions of these
theories without any approximation technique (except for genus expansion).
The distinctive features of our approach are both its effectiveness -- we
are able to compute explicit solutions -- and the method itself -- we do
not use any continuum limit tecnique. This allows us to find many
solutions which are invisible to the latter technique, as will be clear
in a moment.

The essential tools used in this paper are the same introduced in the
previous ones. In particular here we rely on the basic analysis
carried out in \cite{BX2}, although the emphasis on particular
aspects may have changed in the meantime.

For a model of $N\times N$ hermitean matrices with bilinear
coupling, we proceed as
follows. 1) We associate to the initial matrix model
partition function $Z_N$ a discrete linear system (there are several
equivalent ones), we identify the coupling conditions (string equations)
and finally we specify how we can reconstruct $Z_N$ from the quantities
appearing in the linear system; we obtain in this way an
extended description of our matrix model
in the sense that this description is equivalent to the original path integral
whenever the latter exists, but it is valid also for values of the coupling
parameters for which the path integral is not defined. 2) We derive the
$W_\infty$ constraints.

At this point we are faced with an integrable hierarchy
of {\it differential--difference} equations; the solutions we are interested in
are those that satisfy the above $W_\infty$ constraints. To select them out
we continue as follows.

3) We separate the dependence on $N$ from the dependence on the coupling
parameters via a procedure introduced in \cite{BX2}, which consists in
substituting the first flow equations into the remaining ones. This leaves
us with a hierarchy of {\it purely differential} equations. 4) This new
hierarchy is also integrable and contains integrable subsystems (reduced
hierarchies) which can be nicely classified. 5) We identify the critical
points. 6) We integrate the differential equations and plug the solutions
into the $W_\infty$ constraints to fix possible integration constants.
In this way we obtain the sought for correlation functions.

As one will see, all the above steps are of {\it algorithmic} nature.

The results of our approach are of two types. The first, preliminary to
the second, consists in the identification and analysis of the
integrable hierarchy of differential equations which appear in two--matrix
models. Since this is a long and self--consistent analysis we preferred
to publish it in a separate paper \cite{BX5}. The second group of
results is collected in this paper and concerns more specifically
the calculus of correlation functions of two--matrix models.
We are interested in the genus expansion of the solutions and concentrate
in particular on genus 0 results. They can be compared with the 2D gravity
theory in the continuum and with topological field theory models.

There are two broad families of models and correlation functions (CF's).
To understand this point one should remember that two--matrix models describe
the interaction of the first matrix $M_1$ with itself, which is described
by a (possibly infinite) polynomial in $M_1$, of the second matrix $M_2$
with itself, which is described by a similar polynomial in $M_2$, and
finally by a bilinear interaction $M_1M_2$ with a coupling $c$.
The first family of models is the one in which $c\neq 0$, i.e. we
consider the whole theory.
The second family consists of models in which
either $c=0$ or $c$ does not appear, and the $M_2$ sector is disregarded.
We provide one example of the first family, but we mostly concentrate on the
second. In the latter the matrix $M_2$ plays a spectator role; its
self--interaction and interaction with $M_1$ are however essential.
This goes as follows. {\it Every model is characterized by an integrable
hierarchy, which is a reduction of the KP hierarchy}.
The order of the second potential determine a subclass of these models:
if the order is $p$ then the model is characterized by either a $2p$--field
representation of the KP hierarchy or by one of its integrable reductions.
The models are accordingly labeled ${\cal M}_p^l$ with $l=0,\ldots,p-1$,
where ${\cal M}_p^{p-1}$ corresponds to the just mentioned $2p$--field
representation of KP, while the others are reductions. In particular
${\cal M}_p^0$ corresponds to the $p$--th KdV hierarchy. The latter are
characterized by the fact that the correlation functions do not depend on
the size of the matrices, $N$. For all the other models the correlation
functions do depend on $N$, therefore these models cannot be `seen' with
continuum limit techniques.
The first potential has another role, the non--vanishing coupling parameters
in it determine the {\it small phase space} and the critical points.

We remark that an explicit solution of the relevant hierarchy,
satisfying the W--constraints (string equation), underlies the explicit
calculation of the correlation functions of a given model.
The solutions of the hierarchies relevant for CF's are characterized by
a genus by genus homogeneity, defined according to a degree which is determined
by the particular critical point one is considering. This is a distinctive
feature of enormous practical utility.

In elaborating our method we have been inspired by earlier works, in particular
\cite{W1},\cite{DW},\cite{W2} and \cite{Dijkgraafn}. The models we obtain
are very close to the topological field theory models coupled to gravity
studied therein.

This paper is organized as follows. In section 2 we review the relevant results
of \cite{BX3}. In section 3 we show how one can compute CF's directly from
the $W$-constraints; we show an example of CF's with $c\neq 0$. In section 4
we collect the relevant results of \cite{BX4} and \cite{BX5}. Section 5 is
entirely devoted to the explicit calculation of genus 0 CF's at the first
critical point via integration of
the relevant hierarchy in some specific models, precisely the three
models specified by the 4--field representation of the KP--hierarchy and
its integrable reductions. In section 6 we give a general prescription for
calculating CF's and a few more explicit examples. Section 7 is devoted to
higher critical points: we show again how to calculate the corresponding
correlation functions. The last section is devoted to a brief discussion of the
connection of the previous models with topological field theories
and 2D gravity coupled to matter. The Appendices contain explicit expressions
for $W$--constraints, flow equations and CF's for the various models.

\section{Review of previous results}

\setcounter{equation}{0}
\setcounter{footnote}{0}

For simplicity we limit ourselves to two--matrix models.
They are initially defined by the partition function
\a
Z_N(t,c)=\int dM_1dM_2 e^{TrU}\0
\b
where $M_1$ and $M_2$ are Hermitian $N\times N$ matrices and
\a
U=V_1 + V_2 + c M_1 M_2\0
\b
with potentials
\a
V_{\al}=\sum_{r=1}^{p_{\al}}t_{\al,r}M_{\al}^r\,\qquad \al=1,2.\label{V}
\b
The $p_{\al}$'s are finite positive integers.

A clarification is in
order concerning the coupling constants $t_{\al,r}$. Later on we will
let $p_\al \to \infty$, thereby introducing an infinite number of
couplings. Therefore the coupling constants split into
two sets, those appearing in eq.(\ref{V}) (which define the model), and
the remaining ones (which are introduced only for computational purposes).
In terms of ordinary field theory the former are analogous to the
interaction couplings, while the
latter correspond to external currents (coupled to composite operators).
In the case of matrix models we will not
make any formal distinction between them. This will allow us to obtain
very symmetrical and powerful formulas -- the $W$--constraints
for example. Of course the distinction is substantial and will appear
when calculating the correlation functions. These will depend
only on the interaction couplings, while the external ones will be
set to zero once they have done their job.

Let us return to the models defined at the beginning.
Our purpose is to study the their correlation functions (CF's),
i.e. the correlation functions of the operators
\a
{\rm Tr}M_1^k, \quad\quad {\rm Tr}M_2^k, \quad\quad {\rm Tr}(M_1M_2) \0
\b
See the beginning of section 3 for more precise definitions.

The ordinary procedure to calculate the partition function consists of
three steps \cite{BIZ},\cite{IZ2},\cite{M}:
$(i)$ one integrates out the angular parts so that only the
integrations over the eigenvalues are left;
$(ii)$ one introduces the orthogonal polynomials
\a
\xi_n(\lambda_1)=\lambda_1^n+\hbox{lower powers},\qquad\qquad
\eta_n(\lambda_2)=\lambda_2^n+\hbox{lower powers}\0
\b
which satisfy the orthogonality relations
\a
\int  d\lambda_1d\lambda_2\xi_n(\lambda_1)
e^{V_1(\lm_1)+V_2(\lm_2)+c\lm_1\lm_2}
\eta_m(\lambda_2)=h_n(t,c)\delta_{nm}\label{orth1}
\b
$(iii)$, using the orthogonality relation (\ref{orth1}) and the properties
of the Vandermonde determinants, one can easily
calculate the partition function
\a
Z_N(t,c)={\rm const}~N!\prod_{i=0}^{N-1}h_i\label{parti1}
\b
Knowing the partition function means knowing
the coefficients $h_n(t,c)$'s.

The information concerning the latter
can be encoded in a suitable linear system plus some coupling conditions,
together with a relation that allows us to identify $Z_N$.
But before we pass to that we need some convenient notations.
For any matrix $M$, we define
\a
\bigl(\cm\bigl)_{ij}= M_{ij}{{h_j}\over{h_i}},\qquad
{\bar M}_{ij}=M_{ji},\qquad
M_l(j)\equiv M_{j,j-l}.\0
\b
As usual we introduce the natural gradation
\a
deg[E_{ij}] = j -i\0
\b
and, for any given matrix $M$, if all its non--zero elements
have degrees in the interval $[a,b]$, then we will simply
write: $M\in [a,b]$. Moreover $M_+$ will denote the upper triangular
part of $M$ (including the main diagonal), while $M_-=M-M_+$. We will write
\a
{\rm Tr} (M)= \sum_{i=0}^{N-1} M_{ii}\0
\b

Analogously, if $L$ is a pseudodifferential operator, $L_+$ means the purely
differential part of it, while $L_-=L-L_+$.

Let us come now to the step 1) mentioned in the introduction.
First it is convenient to pass from the basis of orthogonal polynomials
to the basis of orthogonal functions
\a
\Psi_n(\lambda_1)=e^{V_1(\lambda_1)}\xi_n(\lambda_1),
\qquad
\Phi_n(\lambda_2)=e^{V_2(\lambda_2)}\eta_n(\lambda_2).\0
\b
The orthogonality relation (\ref{orth1}) becomes
\a
\int d\lm_1 d\lm_2\Psi_n(\lambda_1)e^{c\lm_1\lm_2}
\Phi_m(\lambda_2)=\delta_{nm}h_n(t,c).\label{orth2}
\b
As usual we will denote the semi--infinite column vectors with components
$\Psi_0,\Psi_1,\Psi_2,\ldots,$ and  $\Phi_0,\Phi_1,\Phi_2,\ldots,$
by $\Psi$ and $\Phi$, respectively.

Next we introduce the following $Q$--type matrices
\a
\int d\lm_1 d\lm_2\Psi_n(\lambda_1)
\lm_{\al}e^{c\lm_1\lm_2}
\Phi_m(\lambda_2)\equiv Q_{nm}(\al)h_m=\q_{mn}(\al)h_n,\quad
\al=1,2.\label{Qalpha}
\b
Both $Q(1)$ and $\q(2)$ are Jacobi matrices: their pure upper triangular
part is $I_+=\sum_i E_{i,i+1}$.

Beside the above $Q$ matrices, we will need two $P$--type matrices, defined by
\a
&&\int d\lm_1 d\lm_2\Bigl(\ddlm 1 \Psi_n(\lambda_1)\Bigl)
e^{c\lm_1\lm_2}\Phi_m(\lambda_2)\equiv P_{nm}(1)h_m\\
&&\int  d\lambda_1d\lambda_2\Psi_n(\lambda_1)e^{c\lm_1\lm_2}
\Bigl(\ddlm 2 \Phi_m(\lambda_2)\Bigl)\equiv P_{mn}(2)h_n
\b

The two matrices (\ref{Qalpha}) we
introduced above are not completely independent. More precisely both
$Q(\alpha)$'s can be expressed in terms of only one of them and
one matrix $P$.
Expressing the trivial fact that the integral of the total derivative of the
integrand in eq.(\ref{orth2}) with respect to $\lm_1$ and $\lm_2$
vanishes, we can easily derive the constraints or {\it coupling conditions}
\a
P(1)+c_{12}Q(2)=0,\qquad\quad
c_{12}Q(1)+\bar{\cal P}(2)=0,\0
\b
It is just these coupling conditions that lead to the famous
$W_{1+\infty}$--constraints on the partition function at the discrete level.
We will also refer to them at times as {\it string equations}.
{}From them it follows at once that
\a
Q(\al)\in[-m_{\al}, n_{\al}],\qquad \al=1,2\0
\b
where
\a
m_1=p_2-1, \qquad\quad m_2=1 \0
\b
and
\a
n_1=1, \qquad\quad n_2=p_1-1\0
\b
These equations show that a finite band structure for the
$Q(\alpha)$ is only allowed if the number of perturbations is finite.
Conversely, if we want, say, $Q(1)$ to possess the full discrete KP
structure -- we
should let $p_2\longrightarrow\infty$.

The derivation of the  linear systems associated to our matrix model
is very simple.  We take the derivatives of eqs.(\ref{orth2})
with respect to the time parameters $t_{\al,r}$, and use
eqs.(\ref{Qalpha}).  We get in this way the time evolution of $\Psi$
and $\Phi$, which can be represented in two different ways:

\vskip0.2cm
\noindent
(*)~~~~{$\underline {Discrete~  Linear~ System~~I}$:}
\a
\left\{\ba{ll}
Q(1)\Psi(\lambda_1)=\lambda_1\Psi(\lambda_1),& \\\noal
{\partial\over{\partial t_{1,k}}}\Psi(\lambda_1)=Q^k_+(1)
\Psi(\lambda_1),&1\leq k\leq p_1,\\\noal
{\partial\over{\partial t_{2,k}}}\Psi(\lambda_1)=-Q^k_-(2)
\Psi(\lambda_1),&1\leq k\leq p_2, \\\noal
{\partial\over{\partial\lm}}\Psi(\lambda_1)=P(1)\Psi(\lm_1).&
\ea\right.\label{DLS1}
\b
The corresponding consistency conditions are
\ai
&&[Q(1), ~~P(1)]=1\label{CC11}\\
&&{\partial\over{\partial t_{\al,k}}}Q(1)=[Q(1),~~Q^k_-(\al)]\label{CC12}\\
&&{\partial\over{\partial t_{\al,k}}}P(1)=[P(1),~~Q^k_-(\al)]
\label{CC13}
\bj
(\ref{CC12}) and (\ref{CC13})
are discrete KP hierarchies,
whose integrability and meaning were discussed in \cite{BX3}.

\vskip0.2cm
\noindent(**)~~~~{$\underline {Discrete~  Linear~ System~~II}$:}
\a
\left\{\ba{ll}
\q(2)\Phi(\lambda_2)=\lambda_2\Phi(\lambda_2),\\\noal
{\partial\over{\partial t_{2,k}}}\Phi(\lambda_2)=\q^k_+(2)
\Phi(\lambda_2),& \\\noal
{\partial\over{\partial t_{1,k}}}\Phi(\lambda_1)=-\q^k_-(1)
\Phi(\lambda_2),
&1\leq k\leq p_1\\\noal
{\partial\over{\partial\lm_2}}\Phi(\lambda_2)=P(2)\Phi(\lm_2).&
\ea\right.\label{DLS2}
\b
with consistency conditions
\ai
&&[\q(2),~~P(2)]=1,\label{CC21}\\
&&{\partial\over{\partial t_{\al,k}}}\q(2)=[\q(2),~~\q^k_-(\al)]
\label{CC22}\\
&&{\partial\over{\partial t_{\al,k}}}P(2)=[P(2),~~\q^k_-(\al)]\label{CC23}
\bj

The third element announced in the introduction is the link between the
quantities that appear in the linear system and in the coupling conditions
with the original partition function. We have
\a
{\d \over \d_{\al, r}} \ln Z_N(t,c) = {\rm Tr} \Big(Q^r(\al)), \quad\quad
\al = 1,2 \label{ddZ}
\b
It is evident that, by using the flow equations above we can express all
the derivatives of $Z_N$ in terms of the elements of the $Q$ matrices. For
example
\a
&&{\d^2\over{\d t_{1,1}\d t_{\al,r}}}
\ln Z_N(t,c)=\Bigl(Q^r(\al)\Bigl)_{N,N-1},\label{parti3}\\
&&\qquad\qquad  1\leq r\leq p_{\al};\qquad \al = 1,2\0
\b
 We also recall the coupling dependence of the partition function
\a
{\partial\over{\partial c}}\ln Z_N(t,c)={{\rm Tr}}\Bigl(
Q(1)Q(2)\Bigl)\label{c3}
\b
Knowing all the derivatives with respect to the coupling parameters
we can reconstruct the partition function up to an overall integration
constant.

We also remark that, since the RHS of the above equations is always defined,
they give us a definition of $Z_N$ even in subsets of the parameter space where
the path--integral is ill--defined.

We will be using the following coordinatization of the Jacobi matrices
\a
Q(1)=I_++\sum_i \sum_{l=0}^{m_1} a_l(i)E_{i,i-l}, \qquad\qquad\qquad
\q(2)=I_++\sum_i \sum_{l=0}^{m_2} b_l(i)E_{i,i-l}\label{jacobi}
\b
One can immediately see that
\a
\Bigl(Q_+(1)\Bigl)_{ij}=\delta_{j,i+1}+a_0(i)\delta_{i,j},\qquad
\Bigl(Q_-(2)\Bigl)_{ij}=R_i\delta_{j,i-1}\0
\b
As a consequence of this coordinatization, eq.(\ref{parti3}) gives in
particular the important relation
\a
{\d^2\over{\d t^2_{1,1}}}
\ln Z_N(t,c)=a_1(N),\label{Za1}
\b

To end this subsection we write down explicitly
the $t_{1,1}$-- and $t_{2,1}$--flows, which will play a very important role
in what follows
\ai
&&{\partial\over{\partial t_{1,1}}}a_l(j)=a_{l+1}(j+1)-a_{l+1}(j)
+a_l(j)\Big(a_0(j)-a_0(j-l)\Big)\label{f11}\\
&&{\partial\over{\partial t_{q,1}}}a_l(j)=R_{j-l+1}a_{l-1}(j)-R_ja_{l-1}(j-1)
\label{f21}\\
&&{\partial\over{\partial t_{1,1}}}b_l(j)=R_{j-l+1}b_{l-1}(j)-R_jb_{l-1}(j-1)
\label{f11'}\\
&&{\partial\over{\partial t_{q,1}}}b_l(j)=b_{l+1}(j+1)-b_{l+1}(j)
+b_l(j)\Big(b_0(j)-b_0(j-l)\Big)\label{f21'}
\bj

\subsection{$W_{1+\infty}$ constraints}

The $W_{1+\infty}$ constraints (or simply $W$--constraints)
on the partition function for our two--matrix
model were obtained in \cite{BX3} by putting together both coupling
conditions and consistency conditions (see above). In other words
the $W_{1+\infty}$ constraints contain all the available information.
They take the form
\a
W^{[r]}_n Z_N(t,c)=0, \quad\quad\quad
\tilde W^{[r]}_n Z_N(t,c)=0\quad r\geq0;~~n\geq-r,\label{Wc}
\b
where
\ai
W^{[r]}_n&\equiv& (-c)^n{\cal L}^{[r]}_n(1)-{\cal
L}^{[r+n]}_{-n}(2)\label{Wa}\\
\tilde W^{[r]}_n&\equiv& (-c)^n{\cal L}^{[r]}_n(2)-{\cal L}^{[r+n]}_{-n}(1)
\label{Wb}
\bj

The generators ${\cal L}^{[r]}_n(1)$ are differential operators involving
$N$ and $t_{1,k}$, while ${\cal L}^{[r]}_n(2)$ have the same form
with $t_{1,k}$ replaced by $t_{2,k}$. One of the remarkable
aspects of (\ref{Wc}) is that the dependence on the coupling $c$ is nicely
factorized.
The ${\cal L}^{[r]}_n(1)$ satisfy the following $W_{1+\infty}$ algebra

\ai
&&\relax[{\cal L}^{[1]}_n(1), {\cal L}^{[1]}_m(1)]
 =(n-m){\cal L}^{[1]}_{n+m}(1)\label{LLa}\\
&&\relax[{\cal L}^{[2]}_n(1), {\cal L}^{[1]}_m(1)]
 =(n-2m){\cal L}^{[2]}_{n+m}(1)
 +m(m+1){\cal L}^{[1]}_{n+m}(1)\label{LLb}\\
&&\relax[{\cal L}^{[2]}_n(1),{\cal L}^{[2]}_m(1)]=
 2(n-m){\cal L}^{[3]}_{n+m}(1)-(n-m)(n+m+3){\cal L}^{[2]}_{n+m}(1)
 \label{LLc}
\bj
and in general
\a
[{\cal L}^{[r]}_n(1), {\cal L}^{[s]}_m(1)]=(sn-rm)
 {\cal L}^{[r+s-1]}_{n+m}(1)+\ldots,\label{LLgen}
\b
for $r,s\geq 1;~n\geq-r,m\geq-s$. Here dots denote lower than $r+s-1$ rank
operators. We notice that this $W_{1+\infty}$ algebra is not simple,
as it contains a Virasoro
subalgebra spanned by the ${\cal L}^{[1]}_n(1)$'s. We see that once we
know these generators and ${\cal L}^{[2]}_{-2}(1)$, the remaining ones are
produced by the algebra itself. The explicit expression of the basic generators
is given in Appendix A1.

The algebra of the ${\cal L}^{[r]}_n(2)$ is just a copy of the above one,
and the algebra satisfied by the $W^{[r]}_n$ and by $\tilde W^{[r]}_n$
is isomorphic to both. We refer to this abstract algebra with the symbol
${\cal W}$.

\section{How to compute correlation functions from the $W$--constraints}

\setcounter{equation}{0}
\setcounter{footnote}{0}

This section is a sort of intermezzo. It does not represent our
final solution of the problem. But still it will teach us a lot.
We said above that the $W$--constraints contain all the available information
of our two--matrix model. Therefore, starting from them only, we should be able
to calculate all the correlation functions of the model. This is
indeed true and we are going to give in the following a sample calculation
of this fact. To simplify the notation we set
\a
t_{1,k} \equiv t_k ,&& \quad\quad\quad t_{2,k} \equiv s_k\0\\
{\rm Tr}M_1^k \equiv \tau_k, &&\quad\quad\quad
{\rm Tr}M_2^k \equiv \sigma_k\0
\b
{}From the definition of the partition function we inherit the
definition of the correlation functions
\a
\ll \tau_{k_1}\ldots \tau_{k_n}\gg &=& {\partial\over{\partial t_{k_1}}}
\ldots {\partial\over{\partial t_{k_n}}} \ln Z_N(t,c)\0\\
\ll \sigma_{k_1}\ldots \sigma_{k_n}\gg &=& {\partial\over{\partial s_{k_1}}}
\ldots {\partial\over{\partial s_{k_n}}} \ln Z_N(t,c)\0
\b
These formulae do not make any distinction between internal and external
couplings (see the specification at the beginning of section 2).
Eventually the external couplings will be set to zero.

Now we write the $W$--constraints in this new language. $W^{[1]}_{-1} Z_N=0$
and $\tilde W^{[1]}_{-1} Z_N=0$ become, respectively
\ai
&&\sum_{k=2}^\infty k t_k \ll \tau_{k-1} \gg  +N t_1 +c \ll \tau_1\gg =0
\label{w1-1a}\\
&&\sum_{k=2}^\infty k s_k \ll \sigma_{k-1} \gg  +N s_1 +c \ll \sigma_1\gg =0
\label{w1-1b}
\bj
Instead $W^{[1]}_{0} Z_N=0$ and $\tilde W^{[1]}_{0} Z_N=0$ give rise to
the same equation
\a
\sum_{k=1}^\infty k t_k \ll \tau_k \gg =
\sum_{k=1}^\infty k s_k \ll \sigma_k \gg \label{w10}
\b
The constraint  $W^{[1]}_{1} Z_N=0$ becomes
\ai
&&c \Big(\sum_{l=1}^\infty l t_l \ll \tau_{l+1}\gg +
(N+1) \ll \tau_1\gg \Big)
+ \sum_{l_1, l_2=1}^\infty l_1 l_2 s_{l_1} s_{l_2} \ll \sigma_{l_1 + l_2 -1}
\gg \0\\
&&+\sum_{l=3}^\infty l s_l \sum_{k=1}^{l-2} \Big( \ll \sigma_k \sigma_{l-k-1}
\gg + \ll \sigma_k \gg\ll \sigma_{l-k-1}\gg\Big) \0\\
&&+(2N+1) \sum_{l=2}^\infty l s_l \ll \sigma_{l-1} \gg + (N^2 +N) s_1 =0
\label{w11}
\bj
It should be clear by now that $\tilde W^{[1]}_{1} Z_N=0$ gives an exactly
symmetrical equation where $t_k$ and $\tau_k$ are interchanged with
$s_k$ and $\sigma_k$, respectively. In a similar way we can write
all the other $W$--constraints.

Let us consider now the problem at issue in a simplified situation, i.e.
in genus 0. This will be enough to shed light on the main features.
In this case the $W$--constraints come out simplified, \cite{BIZ}. To see this
assign a suitable degree to the couplings
\a
{\rm deg} \equiv [~~], \quad\quad [t_k] = [s_k] = x-k,\quad [N] = x,
\quad\quad [c] = x-2 \label{deg}
\b
and the degree
\a
[{\cal F}^{(0)}] = 2x\label{degZ}
\b
to the genus zero part of the ``free energy'' (let us set from now on
$\ln Z = {\cal F}$).
Here $x$ is, for the time being, an unspecified positive real number. These
assignments follow directly from the $W$--constraints and the first
flow equations. We have only arbitrarily (but without loss of generality)
fixed the scale. We will be looking for CF's homogeneous (genus by genus)
in the couplings in accordance with the above degree.

In the $W$--constraints the genus 0 contribution is the
homogeneous part of highest degree. In any correlation function we should
henceforth append a label $\ll~\gg _0$ to indicate this contribution.
However we will avoid this cumbersome operation by understanding that
the correlation functions in the remaining part of this section refer only
to the genus 0 contribution. With this convention the genus 0 version
of (\ref{w1-1a}), (\ref{w1-1b}) and (\ref{w10}) remain (formally) the same,
while (\ref{w11}) takes the form
\a
&&c \Big(\sum_{l=1}^\infty l t_l \ll \tau_{l+1}\gg + N \ll \tau_1\gg \Big)
+ \sum_{l_1, l_2=1}^\infty l_1 l_2 s_{l_1} s_{l_2} \ll \sigma_{l_1 + l_2 -1}
\gg \0\\
&+&\sum_{l=3}^\infty l s_l \sum_{k=1}^{l-2}
\ll \sigma_k \gg\ll \sigma_{l-k-1}\gg
+2N \sum_{l=2}^\infty l s_l \ll \sigma_{l-1} \gg + N^2  s_1 =0
\label{w110}
\b
and so on.

The next step consists in specifying the critical point (or the model) we
want to consider. This subject will be explained in detail later. For the
time being we consider the first critical point of the simplest possible
model:
\a
2t_2 = -1 \quad\quad t_k = 0,\quad k>2, \quad\quad\quad
2s_2 = -1 \quad\quad s_k = 0, \quad k>2, \label{cp1}
\b
Therefore CF's will be functions of $N, c, t_1\equiv t$ and $s_1\equiv s$ only.
To distinguish this from the general case we will denote the CF's with
the symbol $<\cdot>$ instead of $\ll \cdot \gg$.
In order to preserve homogeneity, we set
\a
x= 2 \0
\b
in the above degree assignments. The genus 0 $W$--constraints
(\ref{w1-1a}), (\ref{w1-1b}), (\ref{w10}) and (\ref{w110}) become respectively
\ai
&&-<\tau_1> + Nt + c <\sigma_1> =0, \0\\
&&- <\sigma_1 > + N s + c <\tau_1> =0, \0\\
&&- <\tau_2> + t <\tau_1> = - <\sigma_2 > + s <\sigma_1>,\0\\
&& c(- <\tau_3> + t <\tau_2> + N<\tau_1>) \0\\
&&~~~~~~~~~~~~~+ <\sigma_3> - 2 s <\sigma_2>
+(s^2- 2 N) <\sigma_1> + N^2 s =0\0
\bj
Writing down the appropriate formulas for the other constraints one
quickly realizes that these formulas form a recursive and overdetermined
system of algebraic equations  for the one--point correlation functions.
A simple computer program allow us to explicitly calculate them. For
example we obtain
\a
<\tau_1> =N {{t + c s}\over {1- c^2}},\quad\quad
<\sigma_1> = N {{s + c t} \over {1-c^2}}\label{CF0}
\b
and so on (the first few are written down in Appendix C1). We immediately
remark that, setting $c =0$ and $s=0$ in these formulas, we obtain the
CF's of the NLS model already met in the context of one--matrix model,
\cite{BX2}. The NLS model correspond to ${\cal M}_2^1$ in the
classification of the present paper.

One can also derive the multi--point correlation functions by simply
differentiating an appropriate number of times the $W$--constraints  with
respect to the couplings and repeating the same procedure.
However this method of calculating correlation functions is not
very economical since it relies on our being able to solve systems of
algebraic equations which become more and more complicated the higher
the critical point is. There is also another reason why this
method is insufficient: since we do not have at our disposal a theory
of the reduction of the ${\cal W}$ algebra and of the $W$--constraints,
it is impossible to study the reductions of our system, i.e. to study
consistent subsystems which, as we shall see, constitute a large part of
the rich structure of two matrix models.

We need another method and this is
what we are going to study in the following sections. However from the above
example we have extracted some useful information:

1) integrability of the discrete hierarchy on which the $W$--constraints are
based translates itself into a recursive set of equations for the CF's;

2) setting $c=0$ (at the end of the computation) and disregarding
$s_k$ and $\sigma_k$, reduces the CF's to those
of a simpler model, the NLS model studied in \cite{BX2} -- as will be seen,
this is a general fact.

3) it follows from \cite{BX2} that, in the same conditions as at the previous
point, the constraints (\ref{Wc}) can be replaced by the much simpler ones
\a
L_n Z &=&0 \label{effWc}\\
L_n &=& \sum_{k=1}^\infty k t_k \frac{\d}{\d t_{k+n}} + 2N \frac {\d}{\d t_n}
\label{LnNLS}\\
&&~~~~~~~~~~~~~~~~~~~~+\sum_{k=1}^{n-1} \frac {\d^2}{\d t_k \d t_{n-k}}
+ N^2 \delta_{n,0} + Nt_1 \delta_{n,-1}, \quad n\geq -1\0
\b
The $L_n$'s satisfies the Virasoro algebra and these Virasoro constraints
are enough to completely determine the correlation functions.
Effective $W$ constraints of this type will appear also in other cases
(other critical points, other models).

\section{Differential~~
 Hierarchies~~ of~~ Two--Matrix ~~~~ Models.}

\setcounter{equation}{0}
\setcounter{footnote}{0}

The method alluded to at the end of the previous section is based on
extracting differential hierarchies of flow equations from the
discrete ones that characterize two--matrix models. Let us resume the
summary of section 2.
We recalled there that two--matrix models can be represented by means of
coupled discrete linear systems, whose consistency conditions give rise to
discrete KP hierarchies and string equations. Here we review
the method introduced in \cite{BX1} and used in \cite{BX2} and \cite{BX3},
to transform the discrete linear
systems into equivalent differential systems whose consistency conditions are
purely differential hierarchies. This is tantamount to separating
the $N$ dependence from the dependence on the couplings.

The clue to the construction are the first flows, i.e.
the $t_{1,1}$ and $t_{2,1}$ flows. For the sake of simplicity let us
consider the {\it system I} and the flow (\ref{f11}). Let us consider
the generic situation in which $Q(1)$ has $m_1 = p_2-1$ lower diagonal lines
(see the parametrization (\ref{jacobi})). To begin with let us notice that
\a
{\partial\over{\partial t_{1,1}}}\Psi_j=\Psi_{n+1}+a_0(n)\Psi_n\label{t11psi}
\b
and let us adopt
for any function $f(t)$ the convention
\a
f'\equiv {{\partial f}\over{\partial t_{1,1}}}\equiv \partial f,\0
\b
We can rewrite
\a
\Psi_{n}=\hat B_{n}\Psi_{n+1}\label{AB}
\b
where
\a
{\hat B}_n\equiv {1\over{\partial-a_0(n)}}=\d^{-1}\sum_{l=0}^{\infty}
\bigl(a_0(n)\d^{-1}\bigl)^l\label{Bn}
\b
Then it is an easy exercise to prove that the discrete spectral
equation
\a
Q(1)\Psi(\lambda_1)=\lambda_1\Psi(\lambda_1)\0
\b
is transformed into the pseudodifferential one
\a
L_n(1)\Psi_n=\lambda_1\Psi_n\label{spetral1}
\b
where
\a
&&L_n(1)=\d+\sum_{l=1}^{m_1}a_l(n)\hat B_{n-l}\hat B_{n-l+1}\ldots
\hat B_{n-1}\label{Ln}\\
&&\qquad=\d+\sum_{l=1}^{m_1}a_l(n)
{1\over{\d-a_0(n-l)}}\cdot
{1\over{\d-a_0(n-l+1)}}\cdots{1\over{\d-a_0(n-1)}}\0
\b
$L_n(1)$ is an operator of KP type \footnote{In \cite{BX3} we imposed the
conditions $a_l(n)=0, ~n<l$, but actually there is no reason to impose these
conditions on the basis of the matrix model, therefore we drop them here.
Consequently the discussion about non--universality in section 6 of
\cite{BX3} must be dropped too; in the language of that section, the only
alternative present in multi--matrix model is the universal one.}.
Actually it is in general a
reduction of the KP operator
\a
L_{KP}= \d+\sum_{l=1}^{\infty}w_l\d^{-l}\0
\b
where $w_l$ are generic (i.e. unrestricted) coordinates.

Proceeding in the same way for the other equations of {\it system I}
we obtain the new system in differential form
\a
\left\{\ba{ll} L_n(1)\Psi_n=\lambda_1\Psi_n\\\noal
\frac {\d}{ \d t_{1,r}}\Psi_n=\Bigl(L^r_n(1)\Bigl)_+\Psi_n,\\\noal
\frac {\d}{ \d t_{2,r}}\Psi_n=-\Bigl(L^r_n(2)\Bigl)_-\Psi_n,\qquad
\\\noal
M_n(1)\Psi_n=\ddlm 1\Psi_n
\ea\right.\label{diffls1}
\b
In particular we have the following consistency conditions
\a
&&\frac{\partial}{\partial t_{1,r}} L_n(1)=
[ \bigl(L_n^r(1)\bigr)_+, L_n(1)],\label{cflowa}\\
&&\frac{\partial}{\partial t_{2,r}} L_n(1)=[ L_n(1),
\bigl(L_n^r(2)\bigr)_-],\label{cflowb}
\b
They are purely differential equations.

Let us come now to the $n$ dependence of the above equations.
The operator $L_n(1)$ introduced above (\ref{Ln}) depends on the coordinates
of many sectors. We showed in \cite{BX3} that, using the first flow equations,
we can transform it into a pseudodifferential operator depending only on the
$n$--th sector coordinates. This was implemented at the price of
introducing a very complicated expression. For our purposes in this
paper this is not very convenient and we had better proceed in another way.
Precisely we introduce $m_1$ ``fields'' $S_1, ... , S_{m_1}$, related
to the ``field'' $a_0$ in the following way
\a
S_i(n) \equiv a_0 (n-i) \label{Sa0}
\b
Then we can rewrite $L_n(1)$ in the following way
\a
L_n(1)=\d+\sum_{l=1}^{m_1}a_l(n)
{1\over{\d-S_{l}(n)}}\cdot
{1\over{\d-S_{l-1}(n)}}\cdots{1\over{\d-S_1(n)}}\0
\b
So we have achieved the same result as in \cite{BX3}, except that, of course,
the fields $S_i$ are not independent. But following a long tradition in
field theory, we will consider these fields as completely independent from
one another in all the intermediate steps of our calculations and only
eventually impose the condition (\ref{Sa0}).

To further simplify the notation we will consider henceforth the lattice label
$n$ on the same footing as the couplings and write
\a
a_i(n, ...) \equiv a_i (n) ( ...), \quad\quad \quad
S_i(n, ...) \equiv S_i (n)( ...) \0
\b
where dots denote the dependence on $t_{1,k}, t_{2,k}$ and  $c$.
So the expression of $L_1(n)$ get further simplified to
\a
L=\d+\sum_{l=1}^{m}a_l
{1\over{\d-S_{l}}}\cdot
{1\over{\d-S_{l-1}}}\cdots{1\over{\d-S_1}}\label{L}
\b
where, for simplicity, we have dropped the label $(1)$ too.
A similar simplification has to be understood also for the other equations of
the {\it system I} above. This simplified form is the one we constantly
refer to throughout the remaining part of the paper. In particular
we recall the integrable hierarchy
\a
\frac{\partial}{\partial t_{r}} L=
[ \bigl(L^r\bigr)_+, L],\label{cflow}
\b
and $t_r = t_{1,r}$.

We can of course extract an analogous pseudo--differential operator
for {\it system II}, by using the first flows (\ref{f21'}).
One can also reconstruct the discrete systems from the differential ones
and the first flows, for an explicit example see \cite{AFGZ}.

This hierarchy and the relevant integrable reductions have been studied
in \cite{BX4} and in the companion of this paper \cite{BX5}.
There we showed that to each operator such as (\ref{L})
there correspond $m$ distinct integrable reductions which are obtained
via Dirac reduction procedure by successively suppressing the $S$ fields one
by one (the order being irrelevant). Of each of them we gave a Lax
operator representation, which not only constitutes a very quick proof of
integrability, but also provides a very efficient way to explicitly calculate
all flows. To each such Lax operator, as we shall see, we can associate
a full set of correlation functions, i.e. a model. Therefore we can say that
for any $p\equiv p_2= m+1$ ($p_2$ is the order of
the potential $V_2$) there correspond $p$ distinct integrable models
${\cal M}_{p}^l$, where $l$ runs from $p$ to $0$ and counts the number
of $S$ fields. The original unreduced model
${\cal M}_{p}^{p-1}$ is defined of course by the Lax operator (\ref{L}) above.
The most reduced model ${\cal M}_{p}^0$ is characterized by the
$p$--th KdV Lax operator and the $p$--th KdV hierarchy.

To end this section let us comment a bit on the naturalness of the reduction
procedure. Our reduction procedure consists in restricting the second
Hamiltonian according to the chosen constraints and calculating the
second Poisson brackets according to the Dirac recipe. The rest is automatic.
In the framework of the Hamiltonian systems this reduction procedure is
completely natural and allows us to select consistent subsystems
of a given integrable system, or, in other words, to find solutions
of the initial systems in which a part of the degrees of freedom are
disregarded. We stressed in section 2 that our unreduced systems are equivalent
to the initial path integral and that they even provide an enlarged
definition of the latter. However we do not know in general how to go back
from a reduced system to the path integral formulation. We know that
in the simplest case (one--matrix model giving rise to the NLS hierarchy,
see \cite{BX2}) the reduction is the KdV hierarchy and that the KdV
hierarchy corresponds to the even potential case via double scaling limit.
Perhaps a generalization of this fact can be found also for two--matrix models
(see in this sense \cite{DKK}). But we dare say this problem is still open.

\section{The four--field KP hierarchy}

\setcounter{equation}{0}
\setcounter{footnote}{0}

In this section we show how to compute the CF's of a definite model, the
${\cal M}_3^2$, and of its reductions ${\cal M}_3^1$ and ${\cal M}_3^0$,
in genus 0 by explicit use of
the integrable hierarchy. As we shall see, once one knows the hierarchy
of flow equations the CF's are almost completely known. The role
of the $W$--constraints is very limited. Let us start with ${\cal M}_3^2$.

\subsection{The model ${\cal M}_3^2$}

The model ${\cal M}_3^2$ is specified by the Lax operator
\a
L=\d+a_1{1\over{\d-S_1}} + a_2 {1\over{\d-S_2}}{1\over{\d-S_1}}\label{L23}
\b
which arises in the context of the previous section when we consider
a potential $V_2$ of highest order 3. We have therefore altogether four
fields $a_1, a_2, S_1, S_2$. The Poisson brackets, the hamiltonians
and some of the flow equations have been given in \cite{BX4}.

The calculus of CF's via the hierarchy can be subdivided in several steps.

{\it Step 1. Meaning of the fields.}
We specify the connection of the fields characterizing the
model with the correlation functions. Due to equation (\ref{Za1}) we have
\a
a_1 = \ll \tau_1 \tau_1\gg \label{a1}
\b
In the following we will often denote $\tau_1$ by $P$.

The interpretation for $a_2$ can be inferred from the second flow equation for
$a_1$, \cite{BX4}:
\a
\frac{\d a_1}{\d t_2} = \Big(a_1' + 2 a_2 + 2 a_1 S_1\Big)'\label{secflow}
\b
Therefore we have
\a
a_2 = {1 \over 2} \ll \tau_2 P\gg
-{1\over 2} \ll PPP\gg - \ll PP\gg S_1 \label{a2'}
\b
via formal integration of eq.(\ref{secflow}).
Formal integration is an operation we will often use in the following. It is
the formal inverse of $\d$ (no integration constant involved) and is a proper
operation in the context of integrable hierarchies whenever the
pseudodifferential calculus applies. For example it is certainly correct
when, as above, we apply it to expressions containing only the abstract symbols
of the fields, it may be incorrect if applied to a particular solution.
In the latter case one must be careful about the constants of integration.

Going back to eq.(\ref{a2'}), we see that the meaning of $a_2$ in
terms of correlation functions is clear once the
meaning of $S_1$ is clear. An interpretation for the latter and for $S_2$
can be derived from the first flows. In ${\cal M}_3^2$ the first flows
are
\ai
&&a_0'(N) = a_1(N+1) - a_1(N) \0\\
&&a_1'(N)=a_2(N+1) - a_2(N) + a_1(N) \Big( a_0(N) - a_0(N-1)\Big)\0\\
&&a_2'(N) = a_2(N) \Big( a_0(N) - a_0(N-2)\Big)\0
\bj
Let us introduce the operator $D_0= e^{\d_0}-1$ defined by
\a
D_0 f_N = f_{N+1} - f_N\0
\b
for any function $f$ depending on the discrete index $N$.
Using the definition of the $S$ fields,
the first flows can be rewritten as
\ai
&&S_1' + D_0 S_1' = D_0a_1\label{ff1}\\
&&a_1' = D_0 a_2 + a_1 D_0 S_1\label{ff2}\\
&&a_2' = a_2\Big(D_0S_1 + D_0S_2\Big)\label{ff3}
\bj

It is very convenient at this point to consider $t_0\equiv N$ on the same
footing as the couplings $t_k$ and to associate to it an ``operator" $Q$,
in the same way as $t_1$ is associated to $P$, etc.. The label $_0$
used above was motivated by the fact that $\d_0= \frac {\d}{\d t_0}$
mimics the derivative with respect to $N$. Keeping track of
the operator $Q$ is tantamount to studying the $N$ dependence in the CF's.
Then (\ref{ff1}) implies
\a
S_1 = \ll \Big(1- e^{-Q}\Big)P\gg, \quad\quad
S_2 =\ll e^{-Q} \Big(1- e^{-Q}\Big)P\gg\label{S1S2}
\b
Similarly, eqs.(\ref{ff3}) and (\ref{ff2}) become
\ai
&&a_2 = \exp \Big(\ll (e^Q-1)(1-e^{-2Q})\gg\Big) f\label{ff2'}\\
&&D_0 a_2 = \ll PPP\gg - \ll PP\gg \ll (e^Q-1)(1- e^{-Q})P\gg\label{ff3'}
\bj
where $f$ is an integration constant which in general might depend on all the
couplings except $t_1$.
These two are compatibility conditions which come from the request
that $\d $ and $\d_0$ commute.

Eqs.(\ref{a1}, \ref{a2'}) and (\ref{S1S2}) provide the connection between
fields and CF's we were looking for.

{\it Step 2. The genus 0 contribution.} In this
section we will be interested in the genus 0 part of the CF's, which
corresponds
to considering the dispersionless limit of the hierarchy, i.e. the limit
in which, according to the degree analysis in section 3, only the first
derivative terms (in $t_1$ and $t_0$) are retained (see \cite{Kr},\cite{Du}).
This corresponds to assigning the following degrees
\a
\relax [t_k] = x - k~~~ \quad k\geq 0,  \quad\quad\quad [{\cal F}^{(0)}]= 2x \0
\b
as in section 3, and moreover
\a
 [a_1^{(0)}]=2,\quad\quad
[a_2^{(0)}]=3,\quad\quad [S_1^{(0)}] =[S_2^{(0)}] =1\0
\b
and keeping the leading order terms in every equation.
In this limit the two fields
$S_1^{(0)}$ and $S_2^{(0)}$ collapse to a single field $S^{(0)}$ for
\a
S_1^{(0)}= \ll PQ\gg = S_2^{(0)}\0
\b
{}From now on in this subsection we drop the label $^{(0)}$ which indicates the
genus 0 contribution since only the latter will be considered.
With this understanding the first few dispersionless flow equations are
collected in Appendix B1.
The compatibility equations (\ref{ff2'}) and (\ref{ff3'}) become
\ai
&&a_2 = e^{2\ll QQ\gg} f_0(t_0)\label{ff2''}\\
&&\ll PPP \gg = \d_0 a_2 + \ll PP\gg \ll PQQ\gg\label{ff3''}
\bj
Moreover, from (\ref{a2'}),
\a
a_2 = {1 \over 2} \ll \tau_2 P\gg
 - \ll PP\gg \ll PQ\gg \label{a20}
\b

{\it Step 3. The first critical point}. For the type of solutions
we are looking for the first critical point is by definition the one for
which the correlation functions contain only nonnegative integral powers of the
couplings. In this model this circumstance is implemented if
\a
3t_3 = -1, \quad\quad t_k=0 \quad k>3\label{cpS23}
\b
We want to preserve homogeneity, therefore in the above degree assignment
we set $x=3$, i.e.
\a
\relax [t_k] = 3 - k \quad k\geq 0, \quad\quad [a_1]=2,\quad\quad
[a_2]=3,\quad\quad [S] =1, \quad\quad
[{\cal F}^{(0)}]= 6 \label{degS23}
\b
The CF's will therefore be homogeneous functions of $N, t_1, t_2$. At times we
will refer to this subset of the parameter space as the ``small phase space".

This choice of the critical point corresponds, in the path integral language,
to considering the potential
\a
V_1 = t_1 \tau_1 + t_2 \tau_2 -{1\over 3} \tau_3\label{intpot32}
\b
as the interacting potential, while the other $\tau_k$ are regarded as
external fields.

{\it Step 4. Integrating the flow equations.} From the homogeneity ansatz
we have
\a
S= a t_2, \quad\quad a_1 = bt_1 + c t_2^2, \quad\quad a_2 = d t_0 + e t_1 t_2
+ f t_2^3\0\b
where $a,b,c,d,e,f$ are numerical constants to be determined. If we plug these
expressions in the $t_2$ dispersionless flow equations
\a
&&\frac{\d a_1}{\d t_2} = \Big(2 a_2 + 2 a_1 S\Big)'\label{2f1}\\
&&\frac{\d a_2}{\d t_2} = 2 a_2' S + 4 a_2 S'\label{2f2}\\
&& \frac{\d S}{\d t_2} = \Big(S^2 + 2 a_1\Big)'\label{2f3}
\b
we find
\a
a=2b,\quad\quad c= 2 b^2, \quad\quad e=0,\quad\quad f=0\0
\b
Then we can write
\a
&&<PP> = {1\over 2} (at_1 + a^2 t_2^2)\0\\
&&<\tau_2 P> = 2d t_0 + a^2 t_1t_2 + a^3 t_2^3\0\\
&&<PQ> = at_2\0
\b
where, as in section 3, the symbol $<\cdot>$ denotes the correlation function
evaluated in the small phase space.
The second equation is obtained from (\ref{2f1}) via formal
integration.
Integrating these equations with respect to $t_1, t_2$ and $t_0$ respectively,
and comparing the results\footnote{The reason why we have to compare the
results of the different integrations, i.e. keep track of the integration
constants, is that the correlation functions we find do not belong to the
space in which $\d^{-1}$ is a formal integral (for example the space of
rapidly decreasing functions). This remark applies to all the
correlation functions calculated in this paper and means that
integrability is effective in a far larger space than the one in which
the formal rules of the pseudodifferential calculus strictly apply.},
we find $2d=a$ and
\a
<P> = {1\over 4} a t_1^2 + a t_0t_2 + {1\over 2} a^2 t_1 t_2^2 + {1\over 4}
a^3 t_2^4 \label{<P>}
\b

After this example it should be clear how to proceed in order to calculate
$<\tau_2>$. We need to know $<\tau_2 P>$, $<\tau_2 Q>$ and $<\tau_2 \tau_2>$.
The first two are obtained by formally integrating the $t_2$ flows
of $a_1$ and $S$. The third is obtained in the following way. We differentiate
eq.(\ref{a20}) with respect to $t_2$ and get
\a
\frac{\d a_2}{\d t_2} = {1\over 2} \ll \tau_2\tau_2 P \gg - a_1 \frac{\d S}
{\d t_2} - \frac {\d a_1 }{\d t_2} S\0
\b
Then we plug eqs.(\ref{2f1}) and (\ref{2f3}) into this equation and obtain
\a
{1\over 2} \ll \tau_2\tau_2 P\gg = \Big( 4 a_2 S + 2 a_1 S^2 + a_1^2\Big)'
\label{ttP}
\b
This can be formally integrated. So altogether we obtain
\a
&&\ll \tau_2 \tau_2 \gg =  8a_2 S + 4 a_1 S^2 + 2 a_1^2 \0\\
&&\ll \tau_2 P \gg = 2 a_2 + 2 a_1 S \0\\
&& \ll \tau_2 Q \gg = S^2 + 2 a_1 \0
\b
Evaluating the RHS's in the small phase space, integrating and comparing
as before, we find
\a
<\tau_2> = a t_0t_1+2 a^2 t_0t_2^2 + a^3 t_1 t_2^3 + {1\over 2} a^2 t_1^2 t_2 +
{1\over 2} a^4 t_2^5 \label{<t2>}
\b
We can continue in this way and calculate the other one--point CF's, the
pattern for the derivation does not change. They
all depend on the constant $a$. This constant cannot be determined  on the
basis
of the flow equations alone. Before passing to the determination of $a$, let us
remark that while the RHS of eq.(\ref{2f2}) is not a derivative w.r.t. $t_1$,
the RHS of (\ref{ttP}) is. This circumstance, crucial for us to be able to
calculate the CF (otherwise we would be left with an
undetermined integration
constant), is typical of all the flows and is rooted in the integrability
of the hierarchy.

{\it Step 5. The $W$--constraints}.
To determine $a$ we need the $W$--constraints.
Let us consider the $W^{[1]}_{-1}Z_N =0$ constraint at $c=0$. It takes the
form
\a
-<\tau_2> + 2 t_2 <\tau_1> + t_0t_1 =0\0
\b
Inserting the above expressions for $<\tau_2>$ and $<\tau_1>$ we find that
the constant $a$ must be equal to 1. This is a general fact:
provided $a=1$, the correlation
functions calculated by integrating the flow equations satisfy all
the $W$--constraints if $c=0$ (and the $t_{2,k}$ couplings are forgotten).
This confirms what we have already found in section 3 concerning the
NLS correlation functions.

A sample of the one--point CF's  is collected
in Appendix C2.
We recall that multi--point correlation functions
involving only the operators $Q, P, \tau_2$ can be obtained from these
results by simply differentiating a suitable number of times $<P>$ and
$<\tau_2>$ w.r.t. $t_0, t_1$ and $t_2$. For the multi--point
CF's involving $\tau_k$
with $k \geq 3$, we have to first differentiate suitably chosen flow
equations w.r.t. suitably chosen flow parameters, insert the flow equations
in the expressions so obtained and proceed as above.

A remark is in order concerning the constant $a$ determined in this section.
The value of this constant $a=1$ is directly connected with the chosen fixed
point $3t_3 = -1$. The correlation function with generic $a$ are
appropriate to the choice $3t_3 = -{1\over a}$.  In other words
$a$ is a normalization constant \footnote{A normalization problem is always
present in this kind of models and, when comparing CF's obtained in different
contexts or with different methods, one should be careful about the
normalization of both the coupling constants and the critical point.}.
Nevertheless the fact remains that once the critical point is fixed the
hierarchy is not enough to completely determine the CF's. We need also
the $W$--constraints.

{\it Step 6. The CF ~~$<Q>$}. This is the only one--point function we have not

calculated yet. From the above results we have
\a
<PQ> = t_2, \quad\quad\quad <\tau_2 Q > = t_1 + 2 t_2^2 \0
\b
Integrating the first equation w.r.t $t_1$ and the second w.r.t. $t_2$
and comparing, we find
\a
<Q> = t_1 t_2 + {2\over 3} t_2^3 + y t_0\label{<Q>}
\b
where $y$ is an undetermined numerical constant. This of course implies in
particular that
$~~<QQ>= y$. Further information could a priori come from the
compatibility eqs.(\ref{ff2''},
\ref{ff3''}). However, inserting $<PPP> = 1/2$ and $<PQQ>=0$, they only
give us
\a
f_0 ={1\over 2} e^{-2y} t_0\0
\b
Since we have used all the information at hand, the number $y$
remains undetermined.

Finally, integrating $<Q>, <P>$ and $<\tau_2>$ w.r.t. $t_0, t_1$ and $t_2$
respectively, and comparing the results we find
\a
{\cal F}^{(0)} \equiv <1> = t_0 t_1 t_2 + {2\over 3} t_0 t_2^3 + {1\over 4}
t_1^2 t_2^2 + {1\over 2} t_1 t_2^3 + {1\over {12}}  t_1^3 + {1\over 2} y t_0^2
\0
\b

This completes our genus 0 analysis of the model ${\cal M}_3^2$.

\subsection{The model ${\cal M}^1_3$.}

The model ${\cal M}_3^1$ is obtained from ${\cal M}_3^2$ via hamiltonian
reduction -- the constraint is $S_1=0$ --, and it is
specified by the Lax operator
\a
L=\d^2+a_1 + a_2 {1\over{\d-S_2}}\label{L13}
\b
We have therefore three fields $a_1, a_2$ and $ S_2\equiv S$. The Poisson
brackets, the hamiltonians
and some of the flow equations have been given in \cite{BX4}.

The connection of the fields $a_1$ and $S_2$ with the correlation functions
is the same as before .
The interpretation for $a_2$ is slightly different from the previous one.
In fact from the second flow equation for
$a_1$, \cite{BX4}:
\a
\frac{\d a_1}{\d t_2} = 2 a_2'\0
\b
we deduce via formal integration
\a
a_2 = {1 \over 2} \ll \tau_2 P\gg
 \label{a2'''}
\b

The first critical point and the degree assignment are the same as in the
previous model.

{}From now on we will be dealing with genus 0 and before we proceed let us
summarize the basic relations. We have
\a
3t_3 = -1, \quad\quad t_k=0 \quad k>3\0
\b
and
\a
\relax [t_k] = 3 - k \quad k\geq 0, \quad\quad [a_1]=2,\quad\quad
[a_2]=3,\quad\quad [S] =1, \quad\quad
[{\cal F}^{(0)}]= 6 \0
\b
Moreover
\a
a_1 = \ll PP\gg,\quad\quad a_2 = {1 \over 2} \ll \tau_2 P\gg, \quad\quad
S=\ll PQ \gg\label{a1a2S}
\b
and again $Q$ is coupled to $t_0\equiv N$.
The CF's will therefore be homogeneous functions of $N, t_1, t_2$.
We have remarked above that this choice of the critical point corresponds
to considering (\ref{intpot32}) as the interacting potential.

As before we now integrate the dispersionless flow equations (the first
few are collected in Appendix B2).
{}From the homogeneity ansatz we have as before
\a
S= a t_2, \quad\quad a_1 = bt_1 + c t_2^2, \quad\quad a_2 = d t_0 + e t_1 t_2
+ f t_2^3\0\b
where $a,b,c,d,e,f$ are constants to be determined. If we plug these
expressions in the $t_2$ dispersionless flow equations
\a
\frac{\d a_1}{\d t_2} = 2 a_2', \quad\quad
\frac{\d a_2}{\d t_2} = 2 \Big(a_2 S\Big)', \quad\quad
\frac{\d S}{\d t_2} = \Big(S^2 +  a_1\Big)'\0
\b
we find
\a
a=b,\quad\quad,c=0  \quad\quad e=0,\quad\quad f=0\0
\b
Then we can write
\a
&&<PP> = at_1 \0\\
&&<\tau_2 P> = 2d t_0 \0\\
&&<PQ> = at_2\0
\b
Integrating these equations with respect to $t_1, t_2$ and $t_0$ respectively,
and comparing the results, we find $2d=a$ and
\a
<P> = {1\over 2} a t_1^2 + a t_0t_2 \label{<P>13}
\b

Proceeding now in the same way as before and, in particular, using the
identification of $a_2$ contained in (\ref{a1a2S}), we find
\a
<\tau_2> = a t_0 t_1 + a^2 t_0 t_2^2\0
\b
as well as all the one--point functions we wish. All of them depend on
the numerical constant $a$.

It would be natural at this point to plug the one--point functions we found
into the $W$--constraints of subsection 2.2, as before. However one quickly
realizes that this gives inconsistent results. {\it Those $W$--constraints
are incompatible with the reduced hierarchy}. This is not surprising.
One should remember the origin of the $W$--constraints. They come
both from the string equations and the flow equations. Since we have
reduced the hierarchy it is evident that we have to change accordingly
also the $W$--constraints. Unfortunately we do not have a general
theory for reducing the ${\cal W}$ algebra and consequently the
$W$--constraints. However we can easily do without it. Remember that
the only piece of information we need is the constant $a$. It is not difficult
to guess the form of the constraints (at least in the case $c=0$) which
are compatible with the one-point CF's we have calculated. Once this
form is determined it will tell us what the constant $a$ is. In fact,
if we consider the case $c=0$ and disregard the $t_{2,k}$ dependence, it is
enough to calculate the Virasoro constraints, $L_n Z=0 $\footnote{The CF's
of the model ${\cal M}_3^2$ do not satisfy Virasoro constraints such as
$L_n Z=0 $. Therefore some models satisfy Virasoro constraints, while others
do not. This is a phenomenological observation we do not have an explanation
for.}.

The way to proceed is as follows. We try the following form for
$ L_{-1}$
\a
L_{-1} = {1\over 2} \sum_{k=3}^\infty kt_k \frac {\d}{\d t_{k-2}}
+ C_{-1} \label{L-1}
\b
Here $C_{-1} $ is a term not containing derivatives w.r.t. the parameters.
Now we impose $ L_{-1}Z=0$, i.e.
\a
-<P> + 2 C_{-1} =0\0
\b
i.e.
\a
C_{-1} = {1\over 4} a t_1^2 + {1\over 2}a t_0t_2 \0
\b
In a similar way we proceed with
\a
L_0= {1\over 2} \sum_{k=1}^\infty kt_k \frac {\d}{\d t_{k}}
+ C_0\0
\b
and with the other generators. We quickly realize that there exists a solution
only if
\a
a=2\0
\b
What we have done is legitimate if the new generators close over an algebra
of the Virasoro type and if the constraints $L_n Z=0$ are satisfied by
the CF's for $a=2$.
These new constraints are then
compatible with the reduced hierarchy.
These requirements are actually true, to the extent we have been able to verify
them. The new
generators are written down in Appendix A2 and the first few correlation
functions are collected in Appendix C3.

\subsection{The model ${\cal M}^0_3$.}

The model ${\cal M}_3^0$ is obtained from ${\cal M}_3^1$ via hamiltonian
reduction -- the constraint being $S_2=0$ --, and it is
specified by the Lax operator
\a
L=\d^3+a_1\d + a_2 \label{L03}
\b
This is the Lax operator which gives rise to the Boussinesq hierarchy.
We have therefore two fields $a_1, a_2$. The Poisson
brackets, the hamiltonians
and some of the flow equations have been given in \cite{BX4}.
In the Boussinesq hierarchy the $t_{3k}$ flows with $k = 1,2,3...$
do not appear \footnote{From the reduction procedure from ${\cal M}^1_3$
it is possible to define $t_3, t_6, ...$ flows in the reduced model,
but they are not integrable.}. It is therefore natural to ignore $t_0\equiv N$.
This is of course consistent only if we find a solution which does not
depend on $t_k$ with $k=0~{\rm mod}3$. This remarkable fact indeed comes true.

The connection of the fields $a_1$ and $a_2$ with the correlation functions
is the same as in the model ${\cal M}^0_3$
\a
a_1 = \ll PP\gg,\quad\quad a_2 = {1 \over 2} \ll \tau_2 P\gg,\label{a1a2}
\b
But the first critical point and the degree assignment are different.
Precisely we find a solution with the desired properties if we set
\a
4t_4 = -1, \quad\quad t_k=0 \quad k>4\0
\b
and
\a
\relax [t_k] = 4 - k \quad k\geq 0, \quad\quad [a_1]=2,\quad\quad
[a_2]=3,\quad\quad [{\cal F}^{(0)}]= 8 \0
\b

We concentrate now on genus 0, as before.
The CF's will therefore be homogeneous functions of $t_1$ and $t_2$.
As before we now integrate the dispersionless flow equations (the first
few can be derived from Appendix B3).
{}From the homogeneity ansatz we have
\a
a_1 = a t_2,\quad\quad a_2 = b t_1\0
\b
where $a$ and $b$ are numerical constants to be determined. Inserting
this into the $t_2$ flow we find $a= 2b$. Using (\ref{a1a2}) we find
\a
<P>= a t_1t_2\0
\b
Similarly from
\a
<\tau_2 P> = at_1,\quad\quad <\tau_2 \tau_2> = - {2\over 3} a^2 t_2^2\0
\b
we obtain
\a
<\tau_2> = {1\over 2} a t_1^2 - {2\over 9} a^2 t_2^3\0
\b
Moreover
\a
<\tau_4> = {1\over 3} a^2 t_1^2 t_2 - {2 \over{27}} a^3 t_2^4\0
\b
and so on. As before we are left with an undetermined numerical constant
$a$ which only the appropriate $W$--constraints can fix.

We proceed as in the previous example to determine the
$W$--constraints appropriate for this model.
For example we guess the following form for ${\cal L}^{[1]}_{-1}$
\ai
{\cal L}_{-1}^{[1]} &=&
{1\over 3} \sum_{k=4}^\infty{}'{} kt_k \frac {\d}{\d t_{k-3}}
+ C_{-1} \label{L-103}\\
{\cal L}_0^{[1]} &=& {1\over 3} \sum_{k=1}^\infty{}'{} kt_k \frac {\d}{\d
t_{k}}
+ C_0 \label{L003}
\bj
where the prime on the summation symbol means that the sum is over all $k>0$
excluding the multiples of 3. Next we impose
\a
{\cal L}_{-1}^{[1]}Z=0\quad {\rm and}\quad {\cal L}_{0}^{[1]}Z=0\0
\b
We find that these conditions imply
\a
a=6, \quad\quad C_{-1} = 2t_1 t_2, \quad\quad C_0 = {1\over 9} \0
\b

A more thorough analysis confirms this and shows that the constraints
compatible with the Boussinesq hierarchy are
\a
{\cal L}_n^{[r]}Z=0, \quad\quad r=1,2 , \quad\quad n\geq -r \label{constB}
\b
The generators ${\cal L}_n^{[r]}$ with $r=1,2$ form a closed quadratic algebra,
the $W_3$ algebra,
see Appendix A3. It corresponds to the version calculated in \cite{Za} with
central
charge equal to 2. The first few one--point correlation functions of this
model are collected in Appendix C4.

Let us stress the distinctive features of the Boussinesq model:

1) the CF's do not depend on $N$;

2) the appropriate ${\cal W}$ algebra closes over ${\cal L}_n^{[1]}$ and
${\cal L}_n^{[2]}$;

3) the coupling constant $c$ does not come into play.

\section{Other models}

\setcounter{equation}{0}
\setcounter{footnote}{0}

\subsection{General recipe}

The three models discussed above describe essentially all the features of the
${\cal M}_p^l$. And, as far as the method is concerned, it remains the same for
all the other models. For all the ${\cal M}_p^l$ with
$1\leq  l\leq p-1$ the first critical point is fixed by the condition
\a
pt_p =-1,\quad\quad t_k =0\quad k>p \label{cpp}
\b
and the degree assignment is
\a
&&\relax [t_k]= p-k, \quad\quad [{\cal F}]= 2 p \0\\
&&\relax [S] = 1 ,\quad\quad [a_1]=2, \ldots, [a_{p-1}] = p\label{degass}
\b
where $S$ denote the genus 0 part of the fields $S_i$, which is common to all
of them. The CF's will be homogeneous functions of $t_0, t_1, \ldots, t_{p-1}$.

For the models ${\cal M}_p^0$ (the p--th KdV models) instead, one must
first of all disregard all the $t_k$ with $k$ a multiple of $p$;
the first critical point is
\a
(p+1) t_{p+1} = -1, \quad\quad t_k=0 \quad\quad k>p+1\label{cpKdV}
\b
and the degree assignment is
\a
&&\relax [t_k]= p+1-k, \quad\quad [{\cal F}]= 2 p+2 \0\\
&&\relax [S] = 1 ,\quad\quad [a_1]=2, \ldots, [a_{p-1}] = p\label{degassKdV}
\b
The CF's will be homogeneous functions of $t_1,\ldots, t_{p-1}$.

In all the cases the method consists in fixing the form of the fields
on the basis of homogeneity; this leaves a few undetermined numerical
constants; inserting the fields into the flow equations determines
them up to an overall constant; in turn this can be fixed
via the appropriate $W$--constraints. The latter take the universal form
(\ref{Wc}) in all the unreduced ${\cal M}_p^{p-1}$ models. In the other
cases they can be constructed on the basis of the compatibility with
the hierarchy characteristic of the model.

\subsection{Other examples}

For completeness we have explicitly worked out two more examples in genus 0,
${\cal M}_4^3$ and ${\cal M}_4^0$. In the former the fields take the form
\a
S=t_3, \quad\quad a_1 = {2\over 3} t_2 + t_3^2, \quad\quad
a_2 = {1\over 3} t_1 + {2\over 3} t_2t_3 + {2\over 3} t_3^3, \quad\quad
a_3 = {1\over 3} t_0\label{M34}
\b
The first few dispersionless flow equations are collected in Appendix B4
and the first few one--point CF's are in Appendix C5.

As for model ${\cal M}_4^0$ we have
\a
a_1 = 12 t_3,\quad\quad a_2 = 8 t_2,\quad\quad a_3 = 4t_1 + 18
t_3^2\label{4KdV}
\b
The first few dispersionless flows and one--point CF's are in Appendix B5 and
C6, respectively. The generators for the appropriate $W$--constraints
are to be found in Appendix A4.

\section{Higher genus contributions}

\setcounter{equation}{0}
\setcounter{footnote}{0}

Once the genus zero correlation functions of a given model are known,
it is an easy task (at least in principle) to calculate the CF's
in higher genus. There are several methods one can adopt. The most
immediate one is to use the $W$--constraints. These are either the original
constraints (\ref{Wc}) or the effective constraints we calculated in the
previous sections. The important point is that, although the latter have been
determined from the dispersionless versions of the relevant hierarchies,
they are valid in general and provide the contributions to any genus. An
example will suffice to illustrate this point.

Let us consider the NLS model, which we met at the end of section 3 and
was studied in detail in \cite{BX2}. Eq.(\ref{effWc}) gives us the
effective Virasoro constraints. We have seen that the genus zero contribution
is represented by the highest order terms with respect to the degree
assignment (\ref{deg}, \ref{degZ}). An inspection of eqs.(\ref{effWc},
\ref{LnNLS}) tells us that the second nontrivial contribution will have
degree $2x$ less than the leading order, the third $4x$ less, etc.
Therefore the free energy has a genus expansion
\a
{\cal F} = \sum_{h=0}^\infty {\cal F}^{(h)},\quad\quad [{\cal F}^{(h)}]=
2x(1-h)\label{genusexp}
\b
Inserting this into (\ref{effWc}) we find
\a
&&\sum_{k=2}^\infty kt_k \ll \tau_{k+n}\gg_h + 2N\ll \tau_n\gg_h
+\sum_{k=1}^{n-1} \Big(\ll \tau_k \tau_{n-k}\gg_{h-1} \0\\
&&+
\sum_{h'=0}^h \ll \tau_k\gg_{h'} \ll\tau_{n-k} \gg_{h-h'}\Big)+
(N^2 \delta_{n,0}+ Nt_1\delta_{n,-1})\delta_{h,0}=0\label{hgexp}
\b
where $\ll \cdot\gg_h$ represents the genus $h$ contribution to the
$\ll \cdot\gg$ CF. Eqs.(\ref{hgexp}) are recursive, therefore it is
easy to turn the crank and calculate the higher genus contributions

to the order we wish from those of genus 0. A few examples at the
critical point $2t_2=-1$ are given in Appendix D.

{}From the structure of the $W$--constraints appearing in two--matrix
models it is easy to see that genus recursiveness is a general
characteristic. Therefore computability of higher genus contributions
is guaranteed. We recall that genus recursiveness is typical of
topological field theories.

Clearly it is possible to calculate higher genus correlation functions
starting from the relevant hierarchies and computing the genus expansion
of them. This is the best method if one wishes compact formulas for the CF's.

\section{Higher critical points}

\setcounter{equation}{0}
\setcounter{footnote}{0}

The first critical point by definition implies a dependence of the basic
fields on the couplings specified by homogeneous polynomials (i.e. with
non--negative integer powers). Higher critical points are characterized
still by a homogenous dependence, but with rational and/or negative
powers of the couplings. For the models of type ${\cal M}_p^l$ with
$l=1,\ldots, p-1$, characterized by a small phase space with parameters
$t_0, t_1, ... , t_{p-1}$, higher critical points are specified by
\a
kt_k = \pm 1, \quad\quad k>p, \quad\quad t_l=0 \quad {\rm for}~~ l>k
\label{hcp1}
\b
and the degree assignment is \footnote{The parameters $t_0, t_1, \ldots,
t_{p-1}$ are the self--coupling parameters of the model,
while $t_p, \ldots, t_{k-1}$
are to be regarded as external couplings. If one wants models in which
all $t_1,t_2, \ldots , t_{k-1}$ are self--interacting
parameters one should look at models of the type ${\cal M}_k^l$.}

\a
&&\relax [t_l]= k-l, \quad\quad, [{\cal F}]= 2 k \0\\
&&\relax [S] = 1 ,\quad\quad [a_1]=2, \ldots, [a_{p-1}] = p\label{hdegass1}
\b

For the models of type ${\cal M}_p^0$ (p--th KdV hierarchy)
characterized by a small phase space with parameters
$t_1, ... , t_{p-1}$, higher critical points are specified by
\a
kt_k = \pm 1, \quad\quad k>p+1,\quad\quad t_l=0 \quad {\rm for}~~ l>k,
\quad\quad k\neq np \label{hcp2}
\b
and the degree assignment is \footnote{The $t_{p+1}, \ldots, t_{k-1}$
are to be regarded as external couplings, see previous footnote.}
\a
&&\relax [t_l]= k-l, \quad\quad, [{\cal F}]= 2 k \0\\
&&[a_1]=2, \ldots, [a_{p-1}] = p\label{hdegass2}
\b
The $\pm$ sign in eqs.(\ref{hcp1}, \ref{hcp2}) has to be chosen in such a way
as to avoid complex coefficients in the small phase space expressions
of the basic fields.

A few examples will suffice. For the model ${\cal M}_3^0$ the second critical
point is fixed by
\a
5t_5 = 1, \quad\quad t_l=0\quad l>5\0
\b
and, for simplicity, we set $t_4=0$.
Consequently the degree assignment is
\a
\relax [t_l] = 5-k,\quad\quad [{\cal F}] = 10 ,\quad\quad [a_1]=2,
\quad\quad[a_2]=3\0
\b
It is convenient at this point to take immediately into account the constraint
${\cal L}_{-1}^{[1]} Z=0$. It reads
\a
<\tau_2> = -6t_1t_2\label{SE03}
\b
Now eq.(\ref{a1a2}) implies $a_2 = -3 t_2$. Using again the second flow
equations (Appendix B3) we also find
\a
a_1 = 3 t_1^{1\over 2}\label{a1S03}
\b
At this point it is easy to calculate the correlation functions following
the method of section 5. A few of them are collected in Appendix C4.

Another example is the second critical point of the model ${\cal M}_4^0$:
\a
6t_6 =- 1, \quad \quad t_5=0, \quad\quad t_l=0\quad l>6\0
\b
Consequently the degree assignment is
\a
\relax [t_l] = 6-k,\quad\quad [{\cal F}] = 12 ,\quad\quad [a_1]=2,
\quad\quad[a_2]=3, \quad\quad [a_3]=4\0
\b
The constraint ${\cal L}_{-1}^{[1]} Z=0$ tells us that
\a
<\tau_2> =8t_2^2 + 12 t_1t_3\label{SE04}
\b
{}From this and the second flow equations we can derive
\a
a_1 = -{4\over 3} t_1 t_3^{-1}, \quad\quad a_2 = 6t_3, \quad\quad
a_3 = {4\over 9} t_1^2 t_3^{-2} + 4 t_2 \label{a1S04}
\b
from which we can easily derive the CF's (see Appendix C6).

We conclude that we can at least in principle (i.e. modulo technical
difficulties) solve the ${\cal M}_p^l$ models at higher critical points
as well.

\subsection{Non--perturbative equations}

This paper privileges genus expanded solutions.
We do not want to get involved here into the problems of non--perturbative
solutions. However it is useful to spend a few words at least on the
way we can extract non--perturbative equations.
We show a few examples. The first is the NLS model (${\cal M}_2^1$ model)
at the second critical point $3t_3=-1, t_2=0$. Differentiating
the $W_{-1}^{[1]}Z=0$ at $c=0$
w.r.t. $t_0$ and $t_1$ and using the second flow equations one finds
\a
R' + 2RS = t_0, \quad\quad -S' + 2R + S^2 = t_1\label{NPNLS1}
\b
where, as usual, we set $a_1=R$. These two equations must be supplemented
with the $t_0$ flows:
\a
\Big(e^{\d_0} -1\Big)R = e^{\d_0}S', \quad\quad \Big(e^{\d_0} -1\Big)S = (\ln
R)' \label{NPNLS2}
\b
The above equations specify the $t_0, t_1$ dependence of $R$ and $S$.

The second example is the second critical point of the ${\cal M}_3^0$ model,
which has been specified above. Differentiating w.r.t. $t_1$ and $t_2$
the constraint ${\cal L}_{-1}^{[1]}$ and using the second flow equations we
obtain
\a
2a_2 - a_1' + 6t_2=0,\quad\quad 4a_2'  - {7\over 3} a_1'' -
{2\over 3} a_1^2 + 6t_1=0\label{NPS031}
\b
These should be supplemented with the (complete) second flow equations
\a
{{\d a_1} \over {\d t_2} } = 2a_2'-a_1'',\quad\quad
{{\d a_2} \over {\d t_2} } = a_2''-{2\over 3} a_1'''- {2\over 3}a_1 a_1'
\label{NPS032}
\b
The four equations above specify the non--perturbative dependence of $a_1, a_2$
on $t_1$ and $t_2$.

An interesting example is the next critical point in the same model,
${\cal M}_3^0$, i.e.
\a
7t_7 = -1, \quad\quad t_4=0, \quad\quad t_l=0, \quad l>7\label{third}
\b
and $t_5$ is an external coupling. Differentiating the constraint
${\cal L}_{-1}^{[1]}Z=0$ with respect to $t_1$ and $t_2$ we obtain
\a
&& -\ll \tau_4 P\gg + 5t_5 \ll \tau_2 P \gg + 6 t_2 =0\0\\
&&-\ll \tau_4 \tau_2\gg + 5t_5 \ll \tau_2 \tau_2 \gg + 6 t_1 =0\0
\b
respectively. Now, using the $t_2$ and $t_4$ flows (see Appendix B3),
we find
\a
&&a_1''' - 2 a_2'' + 2 a_1a_1' -4 a_1a_2 + 15 t_5 (2a_2 - a_1') + 18 t_2 =0
\label{ising1}\\
&&\frac{1}{3} a_1'''' - 4 a_2^2 + 2 a_1a_1'' + a_1'a_1' + 4 a_1' a_2 +
\frac{8}{3} a_1^3 + 15 t_5 (-\frac{1}{2} a_1'' -\frac{2}{3} a_1^2) + 18t_1=0
\0
\b
With the redefinition
\a
a_1 \rightarrow a_1 - \frac{15}{2} t_5,\quad\quad a_2 \rightarrow
a_2- \frac{1}{2} a_1' \0
\b
and a shift of $t_1$, we obtain the same equations derived in ref.\cite{ising}
(see also \cite{DW})
for the Ising model on discretized Riemann surfaces (up to normalization of
the quantities involved): $t_1, t_2$ and $t_5$ are related to the cosmological
constant, the magnetic field and the parameter $T$ of \cite{ising},
respectively.
Eqs.(\ref{ising1}) together with (\ref{NPS032}) determine
the $t_1, t_2$ dependence of $a_1, a_2$.

\section{Interpretation and comments}

\setcounter{equation}{0}
\setcounter{footnote}{0}

Let us briefly introduce in this last section two points that require
a longer and more careful elaboration:
a possible connection of the previous
models with 2D gravity and topological field theories. We will then comment on
some leftover problems.

On the basis of the continuum theory one would expect the chemical potential
(to be assimilated to $a_1$)
of 2D gravity coupled to conformal matter in genus 0, to behave
like
\a
a_1 \sim t^{2\over{p_0+q_0-1}}\label{asbe}
\b
where $t$ is proportional to the cosmological constant, and $p_0,q_0$ are
relatively prime integers. From the above examples and general statements
one can see that the behaviour of the models
${\cal M}_p^l$ at the various critical points can abundantly account
for (\ref{asbe}), provided we
interpret either $t_1$ or $t_0$ as $t$.

Let us come now to the interpretation of the models we have presented
in this paper in terms of topological
field theories coupled to topological gravity. The latter models are
completely specified once one gives the primaries ${\cal O}_\al$,
the metric $\eta_{\al\beta}$ and the fusion coefficients
$C_{\al\beta}{}^\gamma$. As primaries of our models we can take
the fields coupled to the parameters of the small phase space, $\eta$ and $C$
can be derived from the CF's of the primaries at the first critical point.
For example, for the model ${\cal M}_3^1$ the primaries can be chosen
to be: $\tau_0 \equiv Q$, $\tau_1 \equiv P$ and $\tau_2$ and by
inspection of Appendix C3 we find
\a
\eta_{02} = \eta_{11} = \eta_{20} = 2\0
\b
while the remaining $\eta_{\al\beta}$ vanish. For some models, like the
${\cal M}_p^0$ ones, a similar identification can be very easily
carried out (see also \cite{DW}). There is therefore room for an
interpretation in terms of topological field theories. However
further elaboration is required. We defer a complete discussion
on this point as well
as a study of the connection with Landau--Ginsburg theories to a forthcoming
paper.

It remains for us to point out a few problems concerning the completeness
of the proofs and arguments presented in this paper. The first remark
concerns the statements, made in section 5 and 6, about the
agreement between the calculated correlation functions and the
$W$--constraints.
The agreement has been checked up to a remarkable order (seventh, eighth or
even higher order CF) with a computer. In order to produce a rigorous
proof one should be able to find compact formulae for the CF's, analogous to
those we exhibited in \cite{BX2}. These formulae are perhaps not beyond reach.

A second question is: did we study in this paper all the models contained
in the two--matrix theory? We recall that for any small phase space
of the type $t_1, t_2 , \ldots, t_{p-1}$ or $ t_0, t_1, \ldots, t_{p-1}$,
for any $p$,
we presented one or more models. Does two--matrix theory describe
other integrable models with a more peculiar small phase space (such as,
for example, $t_1, t_2, t_4, t_7$)? To answer this question
one has to ascertain whether there exist other reductions beside those
considered in \cite{BX4},\cite{BX5}.

An entirely new problem is posed by the calculation of the correlation
functions with $c\neq 0$, as we did in section 3, but via the integrable
hierarchies. In this case we have to consider two integrable hierarchies,
the first (system I) and the second (system II). The novelty consists in the
fact that we want to know the dependence of the solutions of the first
on the parameters of the second and vice versa, plus the dependence on
$c$. In \cite{BX3} we showed that the corresponding flows commute.

Finally a few words on general multi--matrix models (with bilinear
couplings). The results of
\cite{BX3} compared with those of \cite{BX5} tell us that very little
has to be expected in terms of new integrable hierarchies and new models.
Of course if one wishes to study CF's dependent on all the couplings (see
the previous paragraph) the problem presents new and perhaps interesting
aspects.

\section*{Appendices}

\subsection*{Appendix A1}

The basic ${\cal L}_n^{[r]}(1)$ generators that appear in (\ref{Wc})
are
\a
{\cal L}^{[1]}_n(1)&\equiv& \sum_{k=1}^{\infty}kt_{1,k}
{\partial\over{\partial t_{1,k+n}}}
+(N+{{n+1}\over2}){\partial\over{\partial t_{1,n}}}
+{1\over2}\sum_{k=1}^{n-1}{{\partial^2}\over{\partial t_{1,k}
\partial t_{1,n-k}}}\0\\
&&+N t_{1,1} \delta_{n,-1} + {1\over 2} N(N+1) \delta_{n,0},~~ n\geq -1.\0
\b
and
\a
{\cal L}^{[2]}_n(1)&\equiv&\sum_{l_1,l_2=1}^{\infty}l_1l_2t_{1,l_1}t_{1,l_2}
{\partial\over{\partial t_{1,l_1+l_2+n}}}
+\sum_{l=1}^{\infty}lt_{1,l}\sum_{k=1}^{l+n-1}
{{\partial^2}\over{\partial t_{1,k}\partial t_{1,l+n-k}}}\0\\
&~&+{1\over3}\sum_{l=1}^{n-2}\sum_{k=1}^{n-l-1}
{{\partial^3}\over{\partial t_{1,l}\partial t_{1,k}\partial t_{1,n-l-k}}}\0\\
&~&+(N+{n\over2}+1)\sum_{k=1}^{n-1}\frac{\d^2}{\d t_k \d t_{n-k}}
+(2N+n+2)\sum_{l=1}^\infty lt_l\frac{\d}{\d t_{n+l}}\0\\
&~&+\bigl(N^2+(n+2)N+{1\over 3}(n+1)(n+2)\bigl)
{\partial\over{\partial t_{1,n}}}\0\\
&~&+(2N^2 t_{1,2} + N t_{1,1}^2) \delta_{n, -2}
+ N(N+1) t_{1,1} \delta_{n,-1}\0\\
&~&+{1\over 3} N(N+1)(N+2) \delta_{n,0},
\qquad\qquad n\geq -2.\0
\b
In the above formulas it is understood that when $t_{1,k}$ appear with
$k \leq 0$, the corresponding term is absent. As already noted
the higher rank generators can be derived from the algebra ${\cal W}$
itself.

Actually it remains for us to specify what
the ${\cal L}_n^{[0]}(1)$'s are. These are given by
\a
{\cal L}^{[0]}_n \equiv  {\partial\over{\partial t_{1,n}}}
+ N \delta_{n,0} , \quad\quad\quad  n\geq 0 \0
\b
These generators represent an abelian extension of the ${\cal W}$ algebra,
of which they form an abelian subalgebra. The other generators
behave tensorially with respect to
${\cal L}^{[0]}_n$. However since the latter play a minor role in this
paper, we do not insist on this point.

\subsection*{Appendix A2}

Here are the generators of the Virasoro algebra for the reduced model
${\cal M}_3^1$:
\a
L_{-1}&=& {1\over 2} \sum_{k=3}^\infty kt_k \frac {\d}{\d t_{k-2}}
+{1\over 2}t_1^2 + Nt_2,\0
\b
\a
L_0&=& {1\over 2} \sum_{k=1}^\infty kt_k \frac {\d}{\d t_{k}} +
{3\over 8} N^2 + {1\over {16}},\0
\b
\a
L_n &=& {1\over 2} \sum_{k=1}^\infty kt_k \frac {\d}{\d t_{k+2n}}
+{1\over 4} N(n+2) \frac{\d}{\d t_{2n}}
 + {1\over 8} \sum_{k=1}^{2n-1} \frac {\d_2}{\d t_k \d t_{2n-k}},
\quad\quad n\geq 1 \0
\b

\subsection*{Appendix A3}

The generators of the ${\cal W}$ algebra appropriate for the
${\cal M}_3^0$ (Boussinesq) model are
\a
{\cal L}^{[1]}_n &=& {1\over 3} \sum_{k=1}^\infty kt_k \frac {\d}{\d
t_{k+3n}}
 + {1\over {18}} \sum_{\stackrel {k,l}{k+l=3n}}
\frac {\d^2}{\d t_k \d t_l}
+{1\over 2}\sum_{\stackrel {k,l}{k+l=-3n}}kl t_kt_l +{1\over 9} \delta_{n,0}
\0,\quad
\forall n\0
\b
\a
{\cal L}^{[2]}_n &=& {1\over 9}
\sum_{l_1,l_2=1}^\infty l_1l_2t_{l_1}t_{l_2}
\frac {\d}{\d t_{l_1+l_2+3n}}+
{1\over {27}} \sum_{\stackrel{l,k,j}{ l-k-j =-
3n}} lt_l\frac {\d^2}{\d t_k \d t_j}\0\\
&&+ {1\over {243}}
\sum_{\stackrel{l,k,j}{l+k+j=3n}}
\frac{\d^3}{\d t_l \d t_k \d t_j} +
{1\over 9}
\sum_{\stackrel{l,k,j}{l+k+j=-3n}}
lkj t_lt_kt_j,
\quad  \forall n\0
\b
In these expressions summations are limited to
the terms such that no index involved is either
negative or multiple of 3.
Contrary to the previous cases we have introduced the generators for all
$n$. In this way the above two sets of generators form a closed algebra,
the $W_3$ algebra, \cite{A},\cite{MK},\cite{PR}
\a
\relax [{\cal L}^{[1]}_n, {\cal L}^{[1]}_m] &=& (n-m){\cal L}^{[1]}_{n+m}
+\frac{1}{6} (n^3 -n) \delta_{n+m,0}\0\\
\relax [{\cal L}^{[1]}_n, {\cal L}^{[2]}_m] &=& (2n-m){\cal L}^{[2]}_{n+m}\0\\
\relax [{\cal L}^{[2]}_n, {\cal L}^{[2]}_m] &=& -\frac{1}{54}(n-m) \Big(
(n^2 + m^2 + 4nm) + 3 (n+m) +2\Big){\cal L}^{[1]}_{n+m} \0\\
&&+ \frac{1}{9} (n-m) \Lambda_{n+m} + {1\over {810}} n(n^2-1)(n^2-4)
\delta_{n+m,0}\0
\b
where
\a
\Lambda_n = \sum_{k\leq -1}{\cal L}^{[1]}_k{\cal L}^{[1]}_{n-k}
+ \sum_{k\geq 0}{\cal L}^{[1]}_{n-k}{\cal L}^{[1]}_k\0
\b
This corresponds to the quantum $W_3$ algebra calculated in \cite{Za} when the
central charge is 2.

\subsection*{Appendix A4}

The generators of the ${\cal W}$ algebra appropriate for the
${\cal M}_4^0$ model are
\a
{\cal L}^{[1]}_n &=& {1\over 4} \sum_{k=1}^\infty kt_k \frac {\d}{\d
t_{k+4n}}
 + {1\over {32}} \sum_{k=1}^{4n+1} \frac {\d^2}{\d t_k \d t_{4n-k}}
+{1\over 2}\sum_{k+l=-4n}kl t_kt_l +{5\over {32}} \delta_{n,0}\0,
\quad \forall n\0
\b
\a
{\cal L}^{[2]}_2 &=& {1\over {16}}
\sum_{l_1,l_2=1}^\infty l_1l_2t_{l_1}t_{l_2}
\frac {\d}{\d t_{l_1+l_2-8}}+
{1\over {64}} \sum_{\stackrel{l,j,k}{ l-j-k =8
}} lt_l\sum\frac {\d^2}{\d t_k \d t_j}\0\\
&&+{1\over {12}}
\sum_{\stackrel{l,k,j}{l+k+j=8}}
lkj t_lt_kt_j,\0
\b
In these expressions summations are limited to
the terms such that no index involved is either
negative or multiple of 4.

{}From the above generators we can generate the $W_4$ algebra.

\subsection*{Appendix B1}

Here are the first few equations of motion of the model  ${\cal M}_3^2$
in the dispersionless limit ($t_k = t_{1,k}$):
\a
\frac{\d a_1}{\d t_2} = \Big(2 a_2 + 2 a_1 S\Big)',\quad\quad
\frac{\d a_2}{\d t_2} = 2 a_2' S + 4 a_2 S',\quad\quad
 \frac{\d S}{\d t_2} = \Big(S^2 + 2 a_1\Big)'\0
\b
\a
&&\frac{\d a_1}{\d t_3} = \Big(3 a_1 S^2  + 6  a_2 S +3 a_1^2\Big)'\0\\
&&\frac{\d a_2}{\d t_3} = 3 a_2' S^2 + 6 a_2 (S^2)'+3a_1a_2' + 6a_1'a_2\0\\
&& \frac{\d S}{\d t_3} = \Big(S^3 + 6 a_1S +3a_2\Big)'\0
\b
\a
&&\frac{\d a_1}{\d t_4} = \Big(4 a_1 S^3  + 12 a_2 S^2 +12 a_1^2S +12 a_1 a_2
\Big)'\0\\
&&\frac{\d a_2}{\d t_4} = 4 a_2' S^3 + 8 a_2 (S^3)'+12a_1a_2'S
+ 24a_2(a_1S)'\0\\
&& \frac{\d S}{\d t_4} = \Big(S^4 + 12 a_1S^2 +12a_2S + 6a_1^2\Big)'\0
\b

\subsection*{Appendix B2}

Here are the first few equations of motion of the model  ${\cal M}_3^1$
in the dispersionless limit and with the convention $t_k = t_{1,k}$:
\a
\frac{\d a_1}{\d t_2} = 2 a_2', \quad\quad
\frac{\d a_2}{\d t_2} = 2 \Big(a_2 S\Big)', \quad\quad
 \frac{\d S}{\d t_2} = \Big(S^2 +  a_1\Big)'\0
\b
\a
&&\frac{\d a_1}{\d t_3} = \Big({3\over 4} a_1^2+ 3  a_2 S \Big)'\0\\
&&\frac{\d a_2}{\d t_3} = \Big({3\over 2}a_1a_2 +3a_2 S^2 \Big)'\0\\
&& \frac{\d S}{\d t_3} = \Big(S^3 + {3\over 2} a_1S +{3\over 2}a_2\Big)'\0
\b
\a
&&\frac{\d a_1}{\d t_4} = \Big(4 a_2 S^2 +4 a_1 a_2\Big)'\0\\
&&\frac{\d a_2}{\d t_4} = \Big(4 a_2 S^3 + 2 a_2^2 + 4a_1a_2S \Big)'\0\\
&& \frac{\d S}{\d t_4} = \Big(S^4 + 2 a_1S^2 +4a_2S + a_1^2\Big)'\0
\b

\subsection*{Appendix B3}

Here are the first few (complete) equations of motion of the model
${\cal M}_3^0$ with the convention $t_k = t_{1,k}$. We
disregard the $t_k$ flows with $k=0$ mod $3$:
\a
{{\d a_1} \over {\d t_2} } = 2a_2'-a_1'',\quad\quad
{{\d a_2} \over {\d t_2} } = a_2''-{2\over 3} a_1'''- {2\over 3}a_1 a_1'
\0
\b
\a
\frac{\d a_1}{\d t_4}&=& \Big(-\frac{1}{3}  a_1''' + \frac{2}{3} a_2''
-\frac{2}{3} a_1 a_1' +\frac{4}{3} a_1 a_2\Big)'\0\\
\frac{\d a_2}{\d t_4}&=&\Big( \frac{1}{3}a_2''' -\frac{2}{9} a_1''''
+\frac{2}{3} a_1a_2' + \frac{2}{3}a_2^2 - \frac{2}{3}a_1 a_1'' -
\frac{1}{2}a_1' a_1' - \frac{4}{27} a_1^3\big)'\0
\b

\subsection*{Appendix B4}

Here are the first few equations of motion of the model  ${\cal M}_4^3$
in the dispersionless limit ($t_k = t_{1,k}$):
\a
&&\frac{\d a_1}{\d t_2} = \Big(2 a_2+ 2 a_1S\Big)' , \quad\quad
\frac{\d a_2}{\d t_2} =  2 a_3' + 2 a_2'S + 4 a_2 S', \0\\
&&\frac{\d a_3}{\d t_2} = 2 a_3' S + 6 a_3 S', \quad\quad
\frac{\d S}{\d t_2} = \Big(S^2 + 2 a_1\Big)'\0
\b
\a
&&\frac{\d a_1}{\d t_3} = \Big(3 a_1^2 + 3 a_3 + 3 a_1 S^2 + 6a_2S\Big)'\0\\
&&\frac{\d a_2}{\d t_3} = 3 a_2 S^2 + 6a_2(S^2)' + 6 a_3' S + 9 a_3 S'
+ 3 a_1 a_2' + 6 a_1'a_2\0\\
&&\frac{\d a_3}{\d t_3} = 3a_1 a_3' + 9 a_1'a_3 + 3 a_3' S^2 + 9 a_3 (S^2)'\0\\
&&\frac{\d S}{\d t_3} = \Big(S^3 + 6a_1S + 3a_2\Big)'
\b

\subsection*{Appendix B5}

Here are the first few equations of motion of the model  ${\cal M}_4^0$
in the dispersionless limit and with the convention $t_k = t_{1,k}$. We
disregard the $t_k$ flows with $k= 0$ mod $4$:
\a
\frac{\d a_1}{\d t_2} = 2 a_2' , \quad\quad
\frac{\d a_2}{\d t_2} =  \Big(2 a_3 - {1\over 2}a_1^2\Big)', \quad\quad
\frac{\d a_3}{\d t_2} = -{1\over 2}a_1' a_2 \0
\b
\a
\frac{\d a_1}{\d t_3} = \Big(- {3\over 8} a_1^2 + 3 a_3\Big)',\quad\quad
\frac{\d a_2}{\d t_3} = -{3\over 4}(a_1a_2)',\quad\quad
\frac{\d a_3}{\d t_3} = - {3\over 8} (a_2^2)' + {3\over 4}a_1 a_3'\0
\b
\a
&&\frac{\d a_1}{\d t_5} = \Big({5\over 8} a_2^2 +{5\over 4} a_1 a_3
- {5\over {32}} a_1^3 \Big)'\0\\
&&\frac{\d a_2}{\d t_5} = \Big({5\over 4}a_2a_3-{15\over
{32}}a_1^2a_2\Big)'\0\\
&&\frac{\d a_3}{\d t_5} = {5\over 8} (a_3^2)' +{5\over {32}} a_1^2 a_3'
-{5\over{16}}(a_1a_2)'a_2\0
\b

\subsection*{Appendix C1}

In this Appendix we collect the first few genus 0 one--point CF's
calculated in section 3. Small phase space:
$N, t_1$. Critical point: $2t_2 =-1$.
\a
<\tau_1> &=&N {{t + c s}\over {1- c^2}},\quad\quad
<\sigma_1> = N {{s + c t} \over {1-c^2}}\0\\
<\tau_2>&=& N{{s^2 + t^2 + 2 cst}\over{(1-c^2)^2}} +
{{N^2 - Ns^2}\over{1-c^2}}\0\\
<\tau_3> &=& N {{t + cs}\over{(1-c^2)^3}}\Big(c^2 s^2 + 2 c s t + t^2 + 3N -
3 Nc^2\Big)\0\\
<\tau_4> &=& \frac{N}{(1-c^2)^4} \Big(c^4 s^4 + 4 c^3 s^3 t + 6 c^2 s^2 t^2 +
4cst^3 + t^4 \0\\
&&+6Nc^2 s^2 - 6 N c^4 s^2 + 12 Nc s t - 12 N c^3 s t \0\\
&&+ 6Nt^2 - 6Nc^2t^2 + 2N^2 - 4 N^2c^2 + 2N^2 c^4\Big)\0
\b
and in general
\a
<\sigma_k>(t, s) = <\tau_k>(s, t)\0
\b

\subsection*{Appendix C2}

One-point correlation functions of the model ${\cal M}_3^2$ in genus 0.
Small phase space: $t_0 \equiv N , t_1, t_2$. Critical point:
$3t_3 =-1$.

\a
<Q> &=& t_1 t_2 + {2\over 3} t_2^3 + y t_0,\0
\b
\a
<P> &=& {1\over 4}  t_1^2 +  t_0t_2 + {1\over 2}  t_1 t_2^2 + {1\over 4}
 t_2^4, \0
\b
\a
<\tau_2> &=&  t_0t_1+2  t_0t_2^2 +  t_1 t_2^3 + {1\over 2}  t_1^2 t_2 +
{1\over 2}  t_2^5, \0
\b
\a
<\tau_3> &=&  {1\over 4} t_1^3 + 3 t_0t_1t_2 + {3\over 2} t_1^2 t_2^2 +
{9\over 4} t_1 t_2^4 + {3\over 4} t_0^2 + 4 t_0 t_2^3 + t_2^6, \0
\b
\a
<\tau_4> &=& {3\over 2} t_0t_1^2 t_1^3 t_2 + 9 t_0t_1t_2^2 + 4 t_1^2 t_2^3
+5 t_1 t_2^5 + 3 t_0^2 t_2 + {{17} \over 2} t_0 t_2^4 + 2 t_2^7,\0
\b
\a
<\tau_5> &=& {5\over 2} t_0^2 t_1 + {{15}\over 2} t_0 t_1^2 t_2 +
25 t_0 t_1 t_2^3 +{{15}\over 4} t_1^3 t_2^2 + {{85}\over 8} t_1^2 t_2^4 \0\\
&&+{5\over {16}} t_1^4 + {{45}\over 4} t_1t_2^6 + 10 t_0^2 t_2^2 +
{{37}\over 2} t_0 t_2^5 + {{65}\over{16}}t_2^8\0
\b
In the first equation $y$ represents an undetermined numerical constant.

\subsection*{Appendix C3}

One-point correlation functions of the model ${\cal M}_3^1$ in genus 0.
Small phase space: $t_0 \equiv N , t_1, t_2$. First critical
point: $3t_3 =-1$.

\a
<Q>&=&2t_1t_2 + {4\over 3} t_2^3+ y t_0  ,\quad\quad
<P> ~=~ t_1^2 + 2 t_0 t_2, \quad\quad <\tau_2>~=~ 2t_0t_1 + 4 t_0  t_2^2\0
\b
\a
<\tau_3> &=& t_1^3 + 6t_0t_1t_2 + 8t_0 t_2^3 + {3\over 4} t_0^2\0
\b
\a
<\tau_4> &=& 4t_0t_1^2 + 16 t_0t_1t_2^2 + 4 t_0^2 t_2 + 16 t_0t_2^4\0
\b
\a
<\tau_5> &=& {5\over 4} t_1^4 + {{15}\over 4} t_1 t_0^2 + 15 t_0 t_1^2 t_2 +
40 t_0 t_1 t_2^3 + 15 t_0^2 t_2^2 + 32 t_0 t_2^5\0
\b

\subsection*{Appendix C4}

One-point correlation functions of the Boussinesq model ${\cal M}_3^0$
in genus 0.
Small phase space: $t_1, t_2$. First critical point: $4t_4 =-1$.
\a
<\tau_1> &=& 6t_1t_2, \quad\quad <\tau_2> ~=~ 3t_1^2 -8t_2^3\0\\
\b
\a
<\tau_4> &=& 12 t_1^2t_2 -16 t_2^4,\quad\quad <\tau_5> ~=~
5t_1^3 -40 t_1t_2^3\0
\b
\a
<\tau_7> &=& 28 t_1^3 t_2 -112 t_1 t_2^4, \quad\quad
<\tau_8>~=~ 10 t_1^4 - 160 t_1^2 t_2^3 + {{256}\over 3} t_2^6\0
\b

Here follow the one-point correlation functions of the
same model at the second critical point: $5t_5 =1$.
\a
<\tau_1> &=& 2t_1^{3\over 2} - 3t_2^2, \quad\quad <\tau_2> ~=~ -6t_1t_2\0\\
\b
\a
<\tau_4> &=& -8 t_1^{3\over 2}t_2 +4 t_2^3,\quad\quad <\tau_5> ~=~
-2t_1^{5\over 2} +15 t_1t_2^2\0
\b

\subsection*{Appendix C5}

Genus 0 one-point correlation functions of the model ${\cal M}_4^3$.
Small phase space: $t_0 \equiv N , t_1, t_2, t_3$. First
critical point: $4t_4 =-1$.
\a
<Q> &=& t_1t_3 +{2\over 3} t_2^2 + 3 t_2 t_3^2 + {9\over 4} t_3^4 + y t_0\0
\b
\a
<P> &=& {2\over 3} t_1t_2 + t_1 t_3^2 + t_0 t_3 + {4\over 3} t_2^2 t_3 +
{{10}\over 3} t_2 t_3^3 + 2 t_3^5\0
\b
\a
<\tau_2> &=& {1\over 3} t_1^2 +{8\over 3} t_1t_2t_3 + {{10}\over 3}t_1t_3^3 +
{4\over 3} t_0 t_2 + 3 t_0 t_3^2 + {8\over {27}} t_2^3 \0\\
&&+ {{16}\over 3} t_2^2 t_3^2 + {{34}\over 3}t_2t_3^4 + {{19}\over 3}t_2^6\0
\b
\a
<\tau_3>&=& 10 t_1 t_2 t_3^2 + 10 t_1t_3^4 + t_1^2 t_3 + t_0 t_1 +
{4\over 3} t_1 t_2^2 + 6 t_0 t_2 t_3\0\\
&&+ 9 t_0 t_3^3 + {{32}\over 9} t_2^3 t_3 + {{68}\over 3} t_2^2 t_3^3 +
38 t_2t_3^5 + 19 t_3^7\0
\b

\subsection*{Appendix C6}

Genus 0 one-point correlation functions of the model ${\cal M}_4^0$.
Small phase space: $t_1, t_2, t_3$. First critical point:
$5t_5 =-1$.
\a
<\tau_1>&=& 8 t_2^2 + 12 t_1t_3,\quad\quad <\tau_2> ~=~ 16 t_1 t_2 - 72 t_2
t_3^2\0
\b
\a
<\tau_3> &=& 6 t_1^2- 72 t_2^2 t_3 + 81 t_3^4\0
\b
\a
<\tau_5> &=& 40 t_1 t_2^2 + 30 t_1^2 t_3 - 360 t_2^2 t_3^2 + 243 t_3^5\0
\b

One-point correlation functions of the same model ${\cal M}_4^0$ at the second
critical point: $6t_6 =-1$.
\a
<\tau_1>&=& 12 t_2t_3 -{2\over 3} t_1^2t_3^{-1},\quad\quad
<\tau_2> ~=~ 12 t_1 t_3 + 8 t_2^2\0
\b
\a
<\tau_3> &=& 12 t_1t_2 +{2\over 9} t_1^3 t_3^{-2} -{{27}\over 2} t_3^3\0
\b
\a
<\tau_5> &=& -{5\over {54}} t_1^4  t_3^{-2} -{{10}\over 3} t_1^2 t_2t_3^{-1}
+{{45}\over 2} t_1 t_3^2 + 30t_2^2 t_3\0
\b

\subsection{Appendix D}

In this Appendix we collect results on correlation functions of the
NLS (${\cal M}_2^1$) model. Small phase space: $N,t_1$. First critical point:
$2t_2=-1$.
The genus $h$ contribution is denoted $<\cdot>_h$.

{\it Genus 0}
\a
<\tau_r>_0 = \sum_{\stackrel {k}{2\leq 2k\leq r+2}} \frac {r!}
{(r-2k+2)!(k-1)!k!} N^k t_1^{r-2k+2}\0
\b

{\it Genus 1}

\a
<\tau_1>_1= <\tau_2>_1 =<\tau_3>=0, \quad\quad <\tau_4>= N,\0
\b
\a
<\tau_5>_1 = 5Nt_1 , \quad\quad <\tau_6>_1= 15 Nt_1^2 + 10 N^2,\0
\b
\a
<\tau_7>_1 = 35 N t_1^3 + 70 N^2t_1, \quad\quad
<\tau_8>_1 = 70 N t_1^4 + 280 N^2t_1^2 + 70 N^3\0
\b
\a
<\tau_9>_1 = 126 Nt_1^5 + 840 N^2t_1^3 + 630 N^3t_1,\0
\b
\a
<\tau_{10}>_1= 210 Nt_1^6 + 2100 N^2t_1^4 + 3150 N^3 t_1^2 + 420 N^4\0
\b
\a
\b
\a
<\tau_1 \tau_i>_1 = 0,\quad 1\leq i\leq 4, \quad\quad<\tau_1\tau_5>_1=5N,
\quad\quad <\tau_1 \tau_6>_1= 30 Nt_1\0
\b
\a
<\tau_2\tau_2>_1=<\tau_2\tau_3>_1=0, \quad\quad <\tau_2\tau_4>_1= 4N\0
\b
\a
<\tau_2\tau_5>_1=30Nt_1, \quad\quad <\tau_3\tau_3>_1= 3N\0
\b

{\it Genus 2}

\a
<\tau_i>_2 =0\quad\quad 1\leq i\leq 7, \quad\quad <\tau_8>_2 = 21N,
\quad\quad<\tau_9>_2=189Nt_1\0
\b
\a
<\tau_{10}>_2= 945 Nt_1^2 + 483 N^2,\quad\quad <\tau_{11}>_2=
3465Nt_1^3 + 5313 N^2t_1 \0
\b


\begin{thebibliography}{}

\bibitem{Douglas} M. Douglas, Phys.Lett.{\bf B238} (90) 176.
\bibitem{is} E.Brezin, M.Douglas, V.Kazakov and S.Shenker,
Phys.Lett.{\bf B237} (1990).

             D.J.Gross and A.A.Migdal, Phys.Rev.Lett. {\bf 64} (1990) 717.
\bibitem{ising} C.Crnkovic, P.Ginsparg, and G.Moore, Phys.Lett.{\bf  B237}
            (1990)196.
\bibitem{G} J.Goeree, Nucl.Phys.{\bf B358}(1991)737.
\bibitem{GN} E.Gava and K.Narain, Phys.Lett.{\bf B263}(1991)213.

            E.Martinec, Comm.Math.Phys. {\bf 138B|} (1991) 437.
\bibitem{tada}T.Tada and T.Yamaguchi, Phys.Lett. {\bf B250} (1990) 38.

              T.Tada, Phys.Lett. {\bf B259} (1991)442.

             S.Odake, Phys.Lett. {\bf B269} (1991)300.
\bibitem{DEB} Jan De Boer, Nucl.Phys. {\bf 336B} (1991) 602.
\bibitem{R} S.Ryang, Phys.Rev.{\bf D46} (1992).
\bibitem{AS} C.Ahn and K.Shigemoto, ``Multi--matrix model and 2D Toda
multi--component hierarchy" CLNS 91/1054.
\bibitem{MM} S.Karchev, A.Marshakov, A.Mironov, A.Morozov and A.Zabrodin,
Nucl.Phys. {\bf 366B} (1991) 569.

A.Marshakov, A.Mironov and A.Morozov, Mod.Phys.Lett.{\bf A7}(1992) 1345.
\bibitem{Kostov} I.K.Kostov, Phys.Lett.{\bf B297} (1992) 74.
\bibitem{PR} S.Panda and S.Roy, Phys.Lett. {\bf B296} (1992) 23.
\bibitem{DKK} J.-M.Daul, V.A.Kazakov and I.K.Kostov {\it Rational Theories
of 2d Gravity from the Two--Matrix Model} CERN-TH6834/93.
\bibitem{Stau} M.Staudacher {\it Combinatorial Solutions of the Two--Matrix
model}, RU--92--64.
\bibitem{BX1} L.Bonora and C.S.Xiong, Phys.Lett.{\bf B285}(1992)191.
\bibitem{BX2} L.Bonora and C.S.Xiong, Int.Jour.Mod.Phys.A {\bf 8}(1993)2973.
\bibitem{BX3} L.Bonora and C.S.Xiong, Nucl.Phys.{\bf 405B} (1993) 191.
\bibitem{BX4} L.Bonora and C.S.Xiong, Phys.Lett.{\bf B317}(1993)329.
\bibitem{BMX} L. Bonora, M. Martellini and C. S. Xiong, Nucl.Phys.
{\bf B375}(1992)453.
\bibitem{X1} C. S. Xiong, Phys.Lett.{\bf B279}(1992)341.
\bibitem{BX5} L.Bonora and C.S.Xiong, {\it The $(N,M)$--th KdV hierarchy
and the associated algebras}, SISSA 171/93/EP, BONN--HE--46/93, AS--ITP--49/93
\bibitem{Ko} M.Kontsevich, Commun. Math.Phys.{\bf 147}(1992)1.
\bibitem{W1} E.Witten, Nucl.Phys.{\bf 340B} (1990) 281.
\bibitem{DW} R.Dijkgraaf and E.Witten, Nucl.Phys. {\bf 342B} (1990) 486.
\bibitem{W2} E.Witten, ``Algebraic geometry associated with matrix models
of two dimensional gravity", IASSNS-HEP-91/74.
\bibitem{Dijkgraafn} R. Dijkgraaf, Lecture notes at Carg\`ese
Summer School, 1991.
\bibitem{BIZ} D. Bessis, C. Itzykson, G.Parisi and J.-B. Zuber,
            Comm.Math.Phys. {\bf 59}(1978) 35.

D. Bessis, C. Itzykson, and J.-B. Zuber, Adv. Appl. Math.{\bf 1}(1980) 109.
\bibitem{IZ2} C.Itzykson and B.Zuber, J.Math.Phys. {\bf 21} (1980) 411.
\bibitem{M} M.L.Metha, Comm.Math.Phys.{\bf 79} (1981) 327.

S. Chedha, G. Mahoux and M. Mehta, J. Phys. A: Math.  Gen.  {\bf 14}(1981)579.
\bibitem{AFGZ} H.Aratyn, L.A.Ferreira, J.F.Gomes and A.H.Zimerman, {\it Toda
and Volterra lattice equations from discrete symmetries of KP hierarchies}
IFT--P.041/93, hep--th/9307147.
\bibitem{A} M.Adler, Invent.Math. {\bf 50}(1979)219.
\bibitem{MK} H.P.McKean, Adv.Math.Supp. {\bf 3} (1978) 217.
\bibitem{Za} A.B.Zamolodchikov, Theor.Math.Phys. {\bf 65}(1986) 1205.
\bibitem{Dickey} L. Dickey, {\it Soliton equations and Hamiltonian Systems},
World Scientific, 1991.
\bibitem{Kr} I.Krichever, Comm.Math.Phys.{\bf 143}(1992)415.
\bibitem{Du} B.Dubrovin, Nucl.Phys.{\bf 379B} (1992) 627.

\end{thebibliography}
\end{document}